\documentclass[11pt,a4paper]{article}
\pdfoutput=1
\usepackage[usenames, dvipsnames]{color}
\usepackage{jheppub}
\usepackage{amssymb,amsmath}
\usepackage{url,booktabs}
\usepackage{multirow}
\usepackage[utf8]{inputenc}
\usepackage[T1,T2A]{fontenc}
\usepackage{slashed}

\usepackage{graphicx}        
\usepackage[export]{adjustbox}

\def\be{\begin{equation}}
\def\ee{\end{equation}}
\def\bea{\begin{eqnarray}}
\def\eea{\end{eqnarray}}

\newcommand{\ch}{\chi^{\pm}_1}
\newcommand{\chp}{\chi^{+}_1}
\newcommand{\chl}{\chi^{-}_1}
\newcommand{\no}{\chi^{0}_1}
\newcommand{\met}{E^{miss}_T}

\usepackage{amsmath,amssymb,amsfonts,mathrsfs}
\usepackage{tikz}
\usetikzlibrary{arrows,shapes}
\usetikzlibrary{trees}
\usetikzlibrary{matrix,arrows} 				
\usepackage{pgffor}							

\tikzstyle{decision} = [diamond, draw, fill=blue!20, 
    text width=4.5em, text badly centered, node distance=3cm, inner sep=0pt]
\tikzstyle{block} = [rectangle, draw, fill=blue!20, 
    text width=5em, text centered, rounded corners, minimum height=2em]
\tikzstyle{line} = [draw, -latex']
\tikzstyle{cloud} = [draw, ellipse,fill=red!20, node distance=3cm,
    minimum height=2em]

\usetikzlibrary{decorations.pathmorphing}	
\usetikzlibrary{decorations.markings}
\tikzset{
    vector/.style={decorate, decoration={snake}, draw},
	provector/.style={decorate, decoration={snake,amplitude=2.5pt}, draw},
	antivector/.style={decorate, decoration={snake,amplitude=-2.5pt}, draw},
    fermion/.style={draw=black, postaction={decorate},
        decoration={markings,mark=at position .55 with {\arrow[draw=black]{>}}}},
    fermionbar/.style={draw=black, postaction={decorate},
        decoration={markings,mark=at position .55 with {\arrow[draw=black]{<}}}},
    fermionnoarrow/.style={draw=black},
    gluon/.style={decorate, draw=black,
        decoration={coil,amplitude=4pt, segment length=5pt}},
    scalar/.style={dashed,draw=black, postaction={decorate},
        decoration={markings,mark=at position .55 with {\arrow[draw=black]{>}}}},
    scalarbar/.style={dashed,draw=black, postaction={decorate},
        decoration={markings,mark=at position .55 with {\arrow[draw=black]{<}}}},
    scalarnoarrow/.style={dashed,draw=black},
    electron/.style={draw=black, postaction={decorate},
        decoration={markings,mark=at position .55 with {\arrow[draw=black]{>}}}},
	bigvector/.style={decorate, decoration={snake,amplitude=4pt}, draw},
}\usetikzlibrary{decorations.markings}

\tikzstyle{block} = [draw, rectangle, 
    minimum height=3em, minimum width=6em]
\DeclareUnicodeCharacter{2212}{-}

\usepackage{ulem}

\graphicspath{{Plots/}}
\relax

\title{	Probing Dark Matter  with Disappearing Tracks at the LHC}

\author[a]{Alexander Belyaev,}
\author[b]{Stefan Prestel,}
\author[c]{Felipe Rojas-Abbate,}
\author[d,e]{Jose Zurita}
\affiliation[a]{Rutherford Appleton Laboratory, Didcot, United Kingdom}
\affiliation[b]{Department of Astronomy and Theoretical Physics,
Lund University, S-223 62 Lund, Sweden}
\affiliation[c]{University of Southampton, Southampton, United Kingdom}
\affiliation[d]{Institute for Nuclear Physics (IKP), Karlsruhe Institute of Technology, Hermann-von-Helmholtz-Platz 1, D-76344 Eggenstein-Leopoldshafen, Germany}
\affiliation[e]{Institute for Theoretical Particle Physics (TTP), Karlsruhe Institute of Technology, Engesserstra{\ss}e 7, D-76128 Karlsruhe, Germany} 
\emailAdd{a.belyaev@phys.soton.ac.uk}
\emailAdd{stefan.prestel@thep.lu.se}
\emailAdd{F.Rojas-Abatte@soton.ac.uk}
\emailAdd{jose.zurita@kit.edu}
\abstract{
	Models where dark matter is a part of an electroweak multiplet feature charged particles with macroscopic lifetimes  due to the charged-neutral mass split of the order of pion mass. At the Large Hadron Collider, the ATLAS and CMS experiments will identify these charged particles as disappearing tracks, since they decay into a massive invisible dark matter candidate and a very soft charged Standard-Model particle which fails to pass the reconstruction requirements. 
While ATLAS and CMS  have focused on the supersymmetric versions of these scenarios, 
	we have performed here the reinterpretation of the latest ATLAS disappearing track search for a suite of dark matter multiplets with different spins and electroweak quantum numbers.
	More concretely, we consider the cases of the inert Two Higgs Doublet model (i2HDM), of Minimal Fermion Dark Matter (MFDM) and of Vector Triplet Dark Matter (VTDM). Our procedure is validated by using the same wino and higgsino benchmark models employed by the ATLAS collaboration. We have found that with
	the disappearing track signature one can probe 
	a vast portion of the
	parameter space, 
	well beyond the reach of prompt missing energy searches (notably mono-jets).
	 We provide tables with the upper-limits on the cross-section upper limits, and efficiencies in the lifetime - dark matter mass plane for all the models under consideration. Moreover we make the recasting code employed here publicly available, as part of the LLP Recasting Repository.}
	 
\arxivnumber{2008.abcde}

\begin{document}
\begin{flushright}
TTP20-029 \\
LU-TP-20-45
\end{flushright}
\maketitle
\notoc

\section{Introduction}
\label{sec:intro}

The existence of Dark Matter (DM) has been established beyond any reasonable doubt by several independent cosmological observations. 
So far,  only  the gravitational interaction of DM  has been experimentally confirmed (for a review see~\cite{Bertone:2004pz}). However, its particle nature and properties are still to be elucidated.

If DM is light enough and interacts with Standard Model (SM) particles directly, or via some mediators with a strength beyond the gravitational one, its elusive nature can be detected or constrained in  direct production at colliders. Therefore, the search for DM in High Energy Physics experiments  became one of the primary goals
of the LHC and  future collider experiments (see e.g~\cite{Boveia:2018yeb} and references therein).

The vanilla DM signal at colliders is the  mono-$X$ signature, where $X$ stands for a SM object,  such as jet, Higgs, Z, W, photon, top-quark, etc that recoils against the missing energy from the DM pair. This signature has  limitations due to the large SM background from $Z\to \nu\nu$, as well as, typically, low signal cross section because of the requirement of large enough missing transverse momentum for the DM pair. In particular, if DM is part of a weak multiplet, and there are no light $Z'$ resonances or other new particles mediating the DM decay (which could enhance DM mono-$X$ signal), even the HL-LHC could probe DM mass only up to about 150-300 GeV, as shown e.g for the case of higgsino DM in MSSM~\cite{Schwaller:2013baa,Barducci:2015ffa,ATLAS:2018diz,CMS:2018qsc}. This limitation motivates us to go beyond the mono-$X$ signature.

In this paper we explore instead another signature, ubiquitous in  BSM models with DM being part of a weak multiplet: \emph{disappearing tracks}. In this scenario, the mass split between DM and its charged multiplet partner(s) can be generically small, in the sub-GeV region, as we explain in what follows. 
If the DM sector consists of just an  electroweak (EW) multiplet~\cite{Cirelli:2005uq}, then DM($D$) and its charged partner(s)($D^{+(+ \dots)}$) masses are degenerate at tree level, as required by the gauge invariance. This is clearly unacceptable from a cosmological point of view. However, this degeneracy is broken due to quantum corrections to $D$ and $D^+$ masses from EW gauge bosons, which introduce the mass split  $\Delta M  \equiv M_{D^+}- M_{D} \propto \alpha_W M_W \sin^2(\theta_W/2)$,  irrespectively of the specific quantum numbers ($Y, T_3$ and spin)  with numerical value around 150-200 MeV. The  mass split, which is non-zero at tree-level in case of scalar DM
(due to the quartic couplings with the SM Higgs field), should also be not too large due to perturbative unitarity constraints, 
which are  particularly important for  large (of the order of 1 TeV) DM mass (see e.g. the case of the  inert two Higgs doublet model(i2HDM)~\cite{Belyaev:2016lok}.)
In attempt to go beyond mono-$X$ signature, one should note that theories with sub-GeV $\Delta M$  do not give any visible decays from $D^+$, even exploiting a boost from initial state radiation, a technique which can be useful for larger mass splits $\Delta M\gtrsim $ 2 GeV~\cite{Schwaller:2013baa}. 

Disappearing tracks occur at $\Delta M$  around 150-200 MeV, when $D^+$ becomes long lived\footnote{There is an ongoing intense activity on studies for Long-Lived Particles (LLPs) at the LHC. We refer the reader to~\cite{Curtin:2018mvb} for a review of the theoretical motivations for LLPs and to~\cite{Alimena:2019zri} for an overview of the existing LHC searches.} with a lifetime of the order of nanoseconds. In such a highly compressed scenario  the 
$D^+ \to D \cal{Y^+}$ decay takes place, with $\cal{Y^+}$ being  $\pi^+$ if $\Delta M > m_{\pi}$ or $\ell^+\nu$ if $\Delta M$ is below the pion mass. The  $\cal{Y^+}$ are very soft and typically stopped by the magnetic field of the detector\footnote{For a strategy to reconstruct the final state pion at ATLAS, see~\cite{ATLAS:2019pjd} and for electron-proton colliders see~\cite{Curtin:2017bxr}. }, thus leaving a short-track ($D^+)$ that ``disappears'' into missing transverse energy ($D$). This disappearing track (DT) signature which we explore in this paper
is very powerful in probing DM scenarios which are  compressed with  $\Delta M \simeq m_{\pi}$.

ATLAS~\cite{Aaboud:2017mpt} and CMS~\cite{Sirunyan:2020pjd}  have actively searched for this signature, often driven by the supersymmetric scenarios of ``pure'' higgsino (weak doublet) and winos (weak triplets). 
The high potential of DT in probing  DM masses in i2HDM model far beyond the mono-jet reach was explored  in~\cite{Belyaev:2016lok} for  i2HDM model and 
in~\cite{Mahbubani:2017gjh,Fukuda:2017jmk,Saito:2019rtg} for  MSSM higgsinos. Furthermore, the impact of disappearing tracks has also been studied for other models of dark matter and neutrino masses (see e.g~\cite{Bharucha:2018pfu,Belanger:2018sti,Filimonova:2018qdc,Jana:2019tdm,Chiang:2020rcv}).

In this study  we present a  simple and flexible  recasting procedure based on  the latest DT search by ATLAS~\cite{Aaboud:2017mpt}, using publicly available information on experimental efficiencies and instrumental backgrounds.
To do this, we first  validate our approach by comparing our results for the MSSM wino and MSSM Higgsino scenarios used by ATLAS as benchmark models. 
Then, we apply our validated procedure to minimal models where DM is either a scalar, fermion or vector.
In order to facilitate the reinterpretation for other DM models, we provide our upper-limits on cross sections and efficiencies in the lifetime--DM mass  ($\tau-M_{DM}$) plane in table format.   In addition,  the software developed for this reinterpretation procedure has been included in the public LLP Recasting Repository~\cite{LLPrepo}, together with a small event sample and the corresponding instructions. We expect this material to be useful to other groups for a straightforward reinterpretation of the ATLAS DT study. 

The paper is structured as follows: in Section~\ref{sec:mod}, we  set our DM model landscape which we use in our study for scalar, fermion and vector EW multiplets.  In Section~\ref{sec:recast}, we briefly summarize the current status of disappearing track searches at the LHC and closely follow the study~\cite{Aaboud:2017mpt}. In this section we define in details our recasting procedure and validate it against the published results for the wino AMSB scenario used as a benchmark by both ATLAS and CMS collaborations. In Section~\ref{sec:results}, we present new results for the LHC limits  for scalar, vector and  fermionic dark matter, highlight the impact of disappearing track searches for  dark sector models beyond the default MSSM benchmarks and make our results  available for a straightforward use by other groups.
In Section~\ref{sec:conclusions}, we draw our conclusions. We reserve Appendix~\ref{app:objects} for the definition of the collider objects employed in this work, Appendix~\ref{app:matching} for a comparison between the MLM and CKKW-L matching schemes, and Appendix~\ref{app:limits} for our cross-section upper limits and efficiencies.



\section{Models with Disappearing Track signatures}
\label{sec:mod}
In this section, we take a closer look into the viable scenarios of dark matter from weak multiplets with different spins and giving rise to disappearing track signatures.

In case of simplest models with just one DM EW multiplet
as an addition  to the SM sector,  the tree-level mass of all multiplet components is the same, as required by the gauge invariance. The charged and neutral components of the multiplet however receive different higher-order corrections. For multiplets with zero hypercharge, the mass of the charged particle(s) is  always above $M_{DM}$\cite{Lahanas:1993ib,Pierce:1994ew,Pierce:1996zz,Drees:1996pk,Fritzsche:2002bi,Martin:2005ch,Yamada:2009ve}\footnote{In case of non-zero hypercharge, negatively charged multiplet members could become lighter than DM mass due to the radiative corrections (depending on their charge and the mass), which eventually makes the model unacceptable.}
and the mass split $\Delta M \sim \alpha_W M_W \sin^2(\theta_W/2)$, which is of the order of the pion mass. This is a very important effect -- it provides the neutral DM candidate and makes the charged particle from the multiplet naturally long-lived.
 
 Two important remarks are in order. First, in the case of the simplest model with a {\it scalar} DM multiplet, the scalar potential has to be supplemented with additional terms allowed by gauge invariance. This can provide a non-zero $\Delta M$ even at tree-level. Second, we note that models with non-zero hypercharge should be rescued from  very high DM direct detection (DD) rates (otherwise they would   blatantly contradict  the experimental results~\cite{Aprile:2018dbl}) because of a non-vanishing $DDZ$ DM interaction with $Z$-boson. For fermionic DM,the minimal way to solve this proble is to introduce a Yukawa term which splits  Dirac DM into two Majorana components as we discuss below.
 
The benchmark models we have chosen are minimal consistent DM models with only few parameters, represented by:
a) inert two Higgs doublet model (i2HDM)\cite{Deshpande:1977rw,Ma:2006km,Barbieri:2006dq,LopezHonorez:2006gr}\footnote{This model is known as Inert doublet model, often denoted as IDM,  but here we use   i2HDM acronym which, to our opinion, reflects better the nature of this model.} for for spin zero DM multiplet;
b) minimal fermion DM model (MFDM), where  DM is a part of EW doublet~\cite{Belyaev:2020xxx});
c) minimal Spin-one Isotriplet Dark Matter model
featuring Dark Matter as a part of vector triplet~\cite{Belyaev:2018xpf}.
Further details on these models are given in the subsections below. We would like to stress that while all these models belong to the thermal Dark Matter class, our findings can be applied to more general scenarios, since our results are presented in a model-independent fashion, in terms of production rates in the lifetime-DM-mass plane.

For the sake of brevity we denote $Z_2$-odd particles from   DM multiplet  as  $D$-particles, and refer to the $Z_2$ symmetry as $D$-parity. This notation will allow us to quickly switch between different models.

\subsection{Inert 2-Higgs Doublet model (i2HDM)}
The i2HDM  is a minimalistic extension of the SM 
with a second scalar doublet $\phi_D$ possessing the same quantum numbers as the SM Higgs doublet $\phi_s$ 
but with no direct coupling to fermions (the inert doublet).
The scalar sector of the model  is given by
\begin{equation}
\mathcal{L}_{i2HDM} =
|D_{\mu}\phi_s|^2 + |D_{\mu}\phi_D|^2 -V(\phi_s,\phi_D)
\textrm{,}
\label{eq:lagrangian}
\end{equation}
where   $V$ is the potential with all scalar interactions compatible with the ${Z}_2$ symmetry:
\begin{eqnarray}
V &=& -m_1^2 (\phi_s^{\dagger}\phi_s) - m_2^2 (\phi_D^{\dagger}\phi_D) + \lambda_1 (\phi_s^{\dagger}\phi_s)^2 + \lambda_2 (\phi_D^{\dagger}\phi_D)^2    \nonumber  \\
&+&  \lambda_3(\phi_s^{\dagger}\phi_s)(\phi_D^{\dagger}\phi_D)
+ \lambda_4(\phi_D^{\dagger}\phi_s)(\phi_s^{\dagger}\phi_D)
+ \frac{\lambda_5}{2}\left[(\phi_s^{\dagger}\phi_D)^2 + (\phi_D^{\dagger}\phi_s)^2 \right]\,.\label{eq:potential}
\end{eqnarray}
In the unitary gauge, the SM doublet $\phi_s$ and the inert doublet $\phi_D$  take the form
\begin{equation}
\phi_s=\frac{1}{\sqrt{2}}
\begin{pmatrix}
0\\
v+H
\end{pmatrix},
\qquad
\phi_D= \frac{1}{\sqrt{2}}
\begin{pmatrix}
\sqrt{2}{D^+} \\
D + iD_2
\end{pmatrix},
\end{equation}
where the first, SM-like doublet,  acquires a vacuum expectation value $v$.
After EW Symmetry Breaking (EWSB), the $D$-parity is preserved by the absence of a vacuum expectation value for the second doublet, which forbids direct coupling of any single inert field to the SM fields, and stabilizes the lightest inert boson.
In addition to the SM-like scalar $H$, the model contains a charged $D^+$ and two  neutral $D$ and $D_2$ scalars from inert doublet. 
Following Ref.~\cite{Belyaev:2016lok},  we denote the two neutral inert scalar masses as $M_{D} < M_{D_2}$, so that we can identify $D$ with the DM candidate.

The model can be conveniently described by a five dimensional parameter space\cite{Belyaev:2016lok}
using the following   phenomenologically relevant variables:
\begin{equation}
\label{eq:model-parameters}
M_{D}\,,\quad M_{D_2} > M_{D}\,,\quad M_{D^+} > M_{D}\,, \quad \lambda_2 > 0\,,\quad \lambda_{345} > -2\sqrt{\lambda_1\lambda_2},
\end{equation}
where $M_{D},M_{D_2}$ and $M_{D^+} $ are, respectively, the masses of the two neutral and  charged inert scalars, whereas   $\lambda_{345}=\lambda_3+\lambda_4+\lambda_5$
is the coupling which governs the Higgs-DM interaction vertex $H D D$.
Constraints  on the parameter space have been comprehensively explored in the literature, see e.g~\cite{Ma:2006km,Barbieri:2006dq,LopezHonorez:2006gr,Arina:2009um,Nezri:2009jd,Miao:2010rg,Gustafsson:2012aj,Arhrib:2012ia,Swiezewska:2012eh,Goudelis:2013uca,Arhrib:2013ela,Krawczyk:2013jta,Krawczyk:2013pea,Ilnicka:2015jba,Diaz:2015pyv,Modak:2015uda,Queiroz:2015utg,Garcia-Cely:2015khw,Hashemi:2016wup,Poulose:2016lvz,Alves:2016bib,Datta:2016nfz,inert-1,inert-2,inert-3,inert-4,Belyaev:2016lok}.

The perturbativity requirement sets an upper limit on the absolute values of the
$\lambda_3,\lambda_4,\lambda_5$ coupling, 
which is controlled by the value of the mass split between $M_D, M_{D_2}$ and $M_{D^+}$. For $M_D \sim$  TeV,
this mass split is limited to be below of about a GeV 
 , which in turn provides the condition for LLPs. To summarise, we see that while in this model $\Delta M$ is non-zero at tree level, it is bounded by perturbativity to be relatively low, especially for large DM masses. Hence, a long-lived $D^+$ can naturally appear this model.

\subsection{Minimal Fermion Dark Matter model (MFDM)}
\label{sec:FDM}
In this model,  DM is a fermion EW doublet with non-zero hypercharge. This scenario is reminiscent of the higgsino-bino system of the MSSM, and also of the singlet-doublet model. As previously discussed, one should implement a mechanism to suppress the DM scattering through Z-boson exchange, in order to comply with the DD constraints from the XENON1T experiment~\cite{Aprile:2018dbl}.

The most minimal way to arrange this is to 
introduce a Yukawa interaction for the  EW doublet with the SM Higgs doublet and an additional Majorana singlet fermion $\chi^0_s$,
resulting to the following Lagrangian~\cite{Belyaev:2020xxx}:
\begin{equation}
\mathcal{L}_{MFDM} = \mathcal{L}_{SM} + 
\bar{\psi}(i\slashed{D}- m_\psi)\psi  
+\frac{1}{2}\bar{\chi^0_s}(i\slashed{\partial}- {m_s})\chi^0_s 
-( Y (\bar{\psi}\Phi\chi^0_s)+ h.c.) ,
\label{eq:mfdm}
\end{equation}
where $\Phi$ is the SM Higgs doublet. The DM 
{$SU(2)$ vector-like doublet with hypercharge $Y=1/2$ is defined as}
\begin{equation}
\psi = \begin{pmatrix} \chi^+ \cr \tfrac{1}{\sqrt{2}}\left( \chi^0_1+i\chi^0_2 \right) \,\end{pmatrix}.
\end{equation}
The last term of Eq.(\ref{eq:mfdm}) is the aforementioned Yukawa  interaction, which splits the neutral
Dirac component of the doublet  into 
two Majorana fermions with distinct mass eigenstates 
$\chi^0_1$ and $\chi^0_2$.
We note that the previously studied doublet-singlet model~\cite{Mahbubani:2005pt,Enberg:2007rp,Cohen:2011ec} has four parameters including $two$ Yukawa couplings, distinguishing left- and right-handed interactions of Higgs and DM doublets with Dirac singlet, $\chi^0_s$. In contrast, this model has 
only one Yukawa coupling involving  the Majorana singlet $\chi^0_s$, and therefore has only three free parameters:  $m_\psi, Y$ and  $m_s$.

The Yukawa interaction mixes  $\chi^0_1$ and $\chi^0_s$ while $\chi^+$ and $\chi^0_2$ have the same  mass $m_\psi$ and remain degenerate at tree-level.
This degeneracy is not essential, since $\chi^0_2$ decay is driven by the $\chi^0_2 \to \chi^0_1 Z^{(*)}$ process. The three parameters  $m_\psi, Y$ and  $m_s$  can be traded for
three physical masses:
\begin{equation}
m_{D}, m_\psi\equiv m_{D^+}=m_{D_2}, \ \ \mbox{and} \ \ m_{D_3},
\end{equation}
corresponding to $(D, D_2, D_3)$ mass bases of the neutral DM sector
with  the eventual mass order
\begin{equation}
m_{D_3}>m_{D^+}=m_{D_2}>m_{D}
\end{equation}
This MFDM model, with singlet-doublet dark sector content, can be mapped onto a bino-higgsino MSSM setup, in which all other SUSY particles (including winos) are decoupled.~\footnote{The main difference between the MFDM and the MSSM (DM higgsino case) is that in the latter the Yukawa coupling is the product of weak couplings and the $\tan \beta$ parameter, which is subject to non-trivial constraints from e.g: flavour physics. We note, however, that this coupling affects the direct detection rates through Higgs exchange, but is otherwise irrelevant for the collider phenomenology, as the production cross sections and the kinematic distributions (for the small mass split) are fully determined by the gauge couplings, spin and weak charge of the EW multiplet.}
In this  model,  DM does not interact with the Z-boson,  because $\chi^0_1$ and $\chi^0_2$  mass eigenstates are split, so the only relevant non-vanishing  $Z\chi^0_1\chi^0_2$ vertex would not provide any DM  direct detection rate at tree level. This allows to avoid strong bounds from DM DD search experiments. 
At the same time, this model can naturally provide the right amount of DM abundance via effective  $D-D_3$ or/and $D-D^+$ co-annihilation or/and DM annihilation via Higgs boson exchange.
The  $D-D_3$ mixing angle $\theta$ and the mass split is defined by Yukawa coupling and $m_\psi$, $m_s$ masses:
\begin{align}\label{tan2theta}
\tan{2\theta} & = 
\frac{2Y v}{m_\psi-m_s}.
\end{align}
One can see that if $m_s \gg m_\psi$, then  $D_3$ decouples and $\Delta M$ becomes small, leaving  the  long-lived $D^+$ and Dark Matter $D$ as the only
experimentally accessible degrees of freedom in the spectrum. This limit has a direct one-to-one correspondence with the so-called ``pure higgsino" MSSM scenario, which is the benchmark model used by ATLAS in~\cite{ATL-PHYS-PUB-2017-019}, and where the relic density is saturated for a dark matter mass of $\sim$1.1 TeV.

\subsection{Minimal Vector Triplet Dark Matter model (VTDM)}
\label{sec:VTDM}

The minimal vector triplet DM model supplements the SM with a new massive vector boson in the adjoint representation of $SU(2)_L$.  The resulting $Z_2$ symmetric  Lagrangian can be written as~\cite{Belyaev:2018xpf}:
\begin{eqnarray}\label{eq:Lagrangian}
\mathcal{L}_{VTDM} & = & \mathcal{L}_{SM}-Tr\left\{ D_{\mu}V_{\nu}D^{\mu}V^{\nu}\right\} +Tr\left\{ D_{\mu}V_{\nu}D^{\nu}V^{\mu}\right\} \nonumber \\
&  & -\frac{g^{2}}{2}Tr\left\{  \left[V_{\mu},V_{\nu}\right]\left[V^{\mu},V^{\nu}\right]\right\} \label{eq:LagUnit}\\
&  & -igTr\left\{ W_{\mu\nu}\left[V^{\mu},V^{\nu}\right]\right\} +\tilde{M}^{2}Tr\{V_{\nu}V^{\nu}\}\nonumber \\
&  & +a\left(\Phi^{\dagger}\Phi\right)Tr\{V_{\nu}V^{\nu}\}\nonumber 
\end{eqnarray}
where $D_{\mu}=\partial_{\mu}-ig\left[W_{\mu},\;\right]$
is the usual $SU(2)_{L}$ covariant derivative in the adjoint representation 
and $V^\mu$ represents
the vector DM iso-triplet.
D-parity prevents the new vector boson mixing with the SM gauge bosons after EWSB (which takes place exactly as in the SM). The physical mass of the new vector bosons, $M_{V}$, is given by
\begin{equation}
M_{V}^{2}=\tilde{M}^{2}+\frac{1}{2}av^{2}\label{eq:MassV}
\end{equation}
where $v \sim 246$~GeV is the usual SM Higgs vacuum expectation value.

In this  model the  mass splitting between $V$ and $V^+$ is induced only at the loop-level. In a manner analogous to the fermionic case,  the neutral and charged isotriplet components are degenerate at tree-level, having the same mass $M_V$, as required by the gauge invariance.  However,  radiative EW corrections  induce a $\Delta M$ split, making the neutral boson lighter than the charged ones. For $M_V \gg M_W,M_Z$, this split is given by~\cite{Belyaev:2018xpf}:
\begin{equation}
\Delta M = \frac{5 g_W^2 (M_W-c_W^2 M_Z)  }{32 \pi } \approx 217.3 \,\rm{MeV},
\label{eqn:mass_splitting}
\end{equation}
and a DM mass of the order of $\sim$3 TeV is necessary to achieve a relic abundance consistent with Planck constraints~\cite{Ade:2015xua}.

\subsection{The lifetime of charged LLPs and the effective W-pion mixing}
\label{sec:W-pikt}

The case of small (below 1 GeV) split $\Delta M=M_{D^+}-M_D$ requires special consideration regarding the calculation of the $D^+$ width, and hence its lifetime.
In particular, for  $\Delta M$ just above the pion mass ( $\Delta M \gtrsim m_{\pi^+} \simeq 140~{\rm MeV}$), $D^+$ will dominantly decay into DM and $\pi^+$. This happens because when the $\Delta M \sim m_\pi$, the naive perturbative calculation of $D^+ \to D  W^{+*} \to  D u\bar{d}$ would underestimate the 
width by about one order of the magnitude, and therefore would overestimate
the lifetime of $D^+$ by the same factor (see e.g. \cite{Belyaev:2016lok} for  detailed discussion).
For a proper evaluation of the lifetime (which is crucial for the LLP phenomenology)  one should use the
$W-\pi$ mixing, described by the non-perturbative term
\begin{equation}
\mathcal{L}_{W\pi}  = \frac{g f_{\pi}}{2\sqrt{2}}W_{\mu}^{+}\partial^{\mu}\pi^{-} + 
\mathrm{h.c.}
\label{eq:wpi}
\end{equation}
with 
 $f_{\pi}=130$ MeV being the  pion decay constant.
This mixing leads to the effective $D^+ D \pi^-$
interaction, which one can derive from the 
$D^+D W^-$ gauge term by means of Eq.(~\ref{eq:wpi}).
The Feynman diagram for this interaction is presented in Fig.\ref{fig:wpi} which in terms  of the effective Lagrangian for DM of spin 0, 1/2 and 1 in the momentum space reads as follows:
\begin{figure}[htb]
	\includegraphics[width=0.3\textwidth,center]{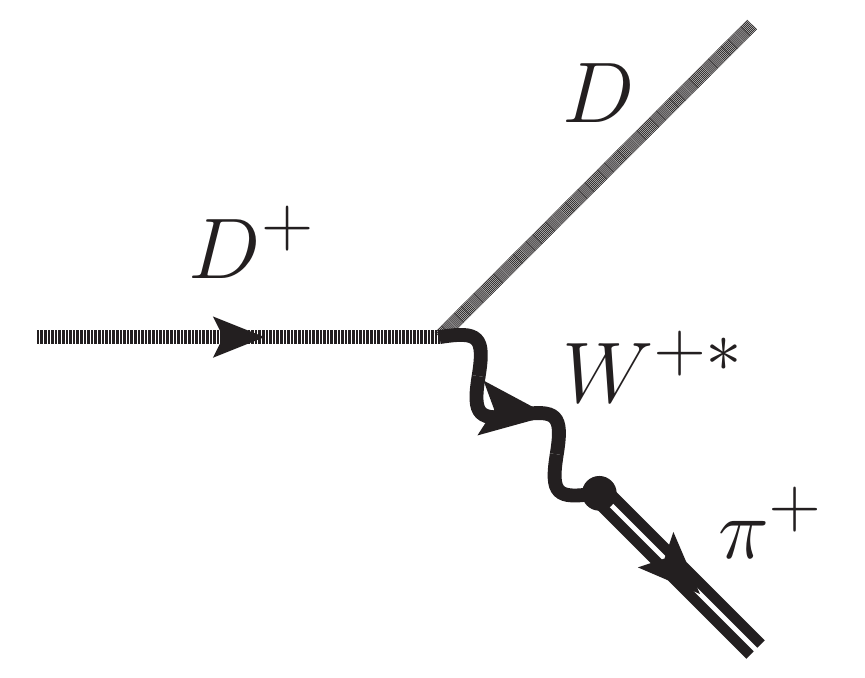}
	\caption{Feynman diagram depicting the  effective 
		$D^+D\pi^-$  interaction from $W-\pi$ mixing.} \label{fig:wpi}
\end{figure}
\begin{eqnarray}
\mathcal{L}_{D^+ D \pi^-}^{\text{i2HDM}} &=& -\frac{g^2f_{\pi}}{4\sqrt{2}M_W^2}
[(p_D-p_D^+)\cdot p_{\pi^-}]
 D^+ D \pi^- + \text{h.c.}
 \label{eq:wpi0.0} \\
\mathcal{L}_{D^+ D \pi^-}^{\text{MFDM}} &=& -\frac{g^2f_{\pi}}{4\sqrt{2}M_W^2}
\cos(\theta_{DD_3})p_{\pi^-}^\mu
D^+\gamma^{\mu}D\pi^-
 + \text{h.c.}  \label{eq:wpi0.5}\\
\mathcal{L}_{D^+ D \pi^-}^{\text{VDM}} &=& -\frac{g^2f_{\pi}}{2\sqrt{2}M_W^2}
\left[
(p_D-p_{D^+})^{\mu} g^{\nu\rho} - 
p_D^{\nu} g^{\mu\rho} + p_{D^+}^{\rho} g^{\mu\nu} \right] {p_{\pi^-}}_\mu D^+_{\nu}D_{\rho}\pi^- + \text{h.c.} \label{eq:wpi1.0} ,
\nonumber
\\
\end{eqnarray}
where $\cos(\theta_{DD_3})$ stands for the cosine of the $D-D_3$ mixing angle for the case of MFDM model. It is worth stressing that the interactions for fermion  DM can have a more general form, by including  different left and right DM couplings. The archetypical example of such a model is the MSSM, where the relevant interactions have the  form
\begin{eqnarray}
\mathcal{L}_{D^+ D \pi^-}^{\text{MSSM}} &=& -\frac{g^2f_{\pi}}{4\sqrt{2}M_W^2}
p_{\pi^-\mu} D^+\left[
g_L \gamma^{\mu} P_L  + 
g_R \gamma^{\mu} P_R  \right] D \pi^-
 + \text{h.c.}, 
\end{eqnarray}
where $g_L$ and $g_R$ are left and right  couplings defined by the specific chargino and neutralino mixings, while  $P_L$ and $P_R$ are the respective  left- and right-handed projectors.

The minimal DM  models discussed above together with the effective $D^+D\pi^-$ interactions given by the  above Eqs.(\ref{eq:wpi0.0}-\ref{eq:wpi1.0}) are implemented into CalcHEP~\cite{Belyaev:2012qa}
using the LanHEP package~\cite{Semenov:1996es,Semenov:1997qm,Semenov:2008jy}, and allows us to effectively and accurately carry out the detailed study  of the LLP  phenomenology of EW DM with different spins, which is presented in the following sections.
We have made these models publicly available at High Energy Physics Model Database (HEPMDB)~\cite{hepmdb}.


\section{Validation of the disappearing track search}
\label{sec:recast}

\subsection{Existing experimental studies}
\label{subsec:ATLAS}

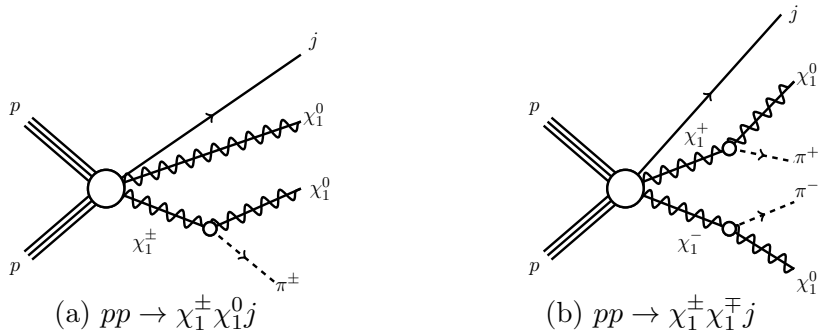
\begin{figure}[ht]
\centering
\resizebox{11cm}{4cm}{
\begin{tikzpicture}[line width=1.3 pt, scale=2.5]
\begin{scope}[shift={(0,0)}]
		\draw[fermionnoarrow] (-0.57,0.53) -- (0.03,0.03);
		\draw[fermionnoarrow] (-0.6,0.5) -- (0,0);
		\draw[fermionnoarrow] (-0.63,0.47) -- (-0.03,-0.03);
			\node at (-0.7,0.6) {$p$};
		\draw[fermionnoarrow] (-0.57,-0.53) -- (0.03,-0.03);
		\draw[fermionnoarrow] (-0.6,-0.5) -- (0,0);
		\draw[fermionnoarrow] (-0.63,-0.47) -- (-0.03,0.03);
			\node at (-0.7,-0.6) {$p$};
		\draw[fermion](0,0) -- (1.5,1);
			\node at (1.6,1.1) {$j$};
		\draw[bigvector](0,0) -- (1.5,0.5);
		\draw[fermionnoarrow](0,0) -- (1.5,0.5);
			\node at (1.6,0.55) {$\no$};
		\draw[fermionnoarrow](0,0) -- (0.8,-0.3);
		\draw[bigvector](0,0) -- (0.8,-0.3);
			\node at (0.3,-0.4) {$\ch$};
		\draw[fermionnoarrow](0.8,-0.3) -- (1.5,0.0);
		\draw[bigvector](0.8,-0.3) -- (1.5,0.0);
			\node at (1.65,0) {$\no$};
		\draw[scalar](0.8,-0.3) -- (1.3,-0.7);
			\node at (1.4,-0.7) {$\pi^{\pm}$};

		\draw[fill=black] (0,0) circle (.14cm);
		\draw[fill=white] (0,0) circle (.14cm);

		\draw[fill=black] (0.8,-0.3) circle (.05cm);
		\draw[fill=white] (0.8,-0.3) circle (.05cm);
 \end{scope}
\begin{scope}[shift={(4,0)}]
	\draw[fermionnoarrow] (-0.57,0.53) -- (0.03,0.03);
		\draw[fermionnoarrow] (-0.6,0.5) -- (0,0);
		\draw[fermionnoarrow] (-0.63,0.47) -- (-0.03,-0.03);
			\node at (-0.7,0.6) {$p$};
		\draw[fermionnoarrow] (-0.57,-0.53) -- (0.03,-0.03);
		\draw[fermionnoarrow] (-0.6,-0.5) -- (0,0);
		\draw[fermionnoarrow] (-0.63,-0.47) -- (-0.03,0.03);
			\node at (-0.7,-0.6) {$p$};
		\draw[fermion](0,0) -- (1.2,1.3);
			\node at (1.3,1.3) {$j$};
		\draw[bigvector](0,0) -- (0.8,0.3);
		\draw[fermionnoarrow](0,0) -- (0.8,0.3);
			\node at (0.55,0.4) {$\chp$};
		\draw[bigvector](0.8,0.3) -- (1.3,0.8);
		\draw[fermionnoarrow](0.8,0.3) -- (1.3,0.8);
		\draw[scalar](0.8,0.3) -- (1.3,0.2);
			\node at (1.4,0.25) {$\pi^{+}$};
		\draw[fermionnoarrow](0,0) -- (0.8,-0.3);
		\draw[bigvector](0,0) -- (0.8,-0.3);
			\node at (0.5,-0.4) {$\chl$};
			\node at (1.4,0.85) {$\no$};
		\draw[fermionnoarrow](0.8,-0.3) -- (1.3,-0.6);
		\draw[bigvector](0.8,-0.3) -- (1.3,-0.6);
			\node at (1.4,0) {$\pi^{-}$};
		\draw[scalar](0.8,-0.3) -- (1.3,-0.1);  //
			\node at (1.4,-0.7) {$\no$};

		\draw[fill=black] (0,0) circle (.14cm);
		\draw[fill=white] (0,0) circle (.14cm);

 		\draw[fill=black] (0.8,-0.3) circle (.05cm);
		\draw[fill=white] (0.8,-0.3) circle (.05cm);
		\draw[fill=black] (0.8,0.3) circle (.05cm);
		\draw[fill=white] (0.8,0.3) circle (.05cm);
 \end{scope}
\end{tikzpicture}
}
\vskip -0.2cm\hspace*{-0.25cm}(a) $pp\rightarrow \ch\no j$\hspace*{0.25\textwidth}(b) $pp\rightarrow \ch\chi^{\mp}_1 j$
\caption{Example diagrams of the signal process used in the analysis.} \label{fig:ewprod}
\end{figure}

Here we will review the most salient features of the ATLAS disappearing track analysis~\cite{Aaboud:2017mpt}. 
The relevant signal process for the models under study  is the pair production of new fields 
are  $pp\rightarrow D^\pm D$ and $pp \rightarrow D^+D^-$.
To trigger events, one can use the initial-state radiation-induced monojet signature, since disappearing tracks contribute to the missing transverse energy (MET)  if the lifetime of $D^{\pm}$ is not too large, such that the $D^{\pm}$ particles do not enter the hadron calorimeter. One should require MET as low as possible for this triggering to keep as many signal event as possible. That is why $pp\rightarrow \ch \no j$ and $pp \rightarrow \chp\chl j$ SUSY processes were the subject of the particular ATLAS study mentioned above. Exemplary Feynman diagrams are shown in figure~\ref{fig:ewprod}.
The event preselection is qualitatively simple and requires the presence of at least one isolated \emph{tracklet}, a large amount of missing transverse energy $\met$ and at least a high $p_T$ jet. A tracklet is a special type of shorter track introduced specifically for this search, and serves as a proxy for the disappearing track signal. Trackets are reconstructed using information from the ATLAS pixel layers of the inner detector, while all other collider objects (jets, muons, electrons) use standard definitions. For completeness we specify the necessary quality cuts on the objects in Appendix ~\ref{app:objects}.

After the preselection stage two more steps follow. First, the {\it Event Selection} takes place, with the goal of isolating the signal from the SM backgrounds. Later, a {\it Tracklet Selection} is carried out. Only good quality tracklets are selected. The ATLAS collaboration has provided in their auxiliary material in HEPDATA\cite{1641262} information to reinterpret (recast) this study. Its proper use requires also to define Generator Level instances of both the {\it Event} and {\it Tracklet} selections, which are obviously based on reconstructed objects. Using public information one can account with a reasonable precision for detector effects for standard objects such as jets, muons, electrons, etc. However, the vital ingredient here is how the parton-level chargino becomes a tracklet (reconstructed-level object). The information at the generator level  provides the recaster a sanity check of the different selections, unfolding reconstruction effects. The combination of the reconstructed and generator level information allows to define a model-independent probability for a parton level chargino to become a tracklet. Our goals here are to first validate the reported probabilities using the same signal model (and parameter points) as in the ATLAS study, and second to apply these validated efficiencies to {a wide class of models under study}.

\subsubsection{The {\it Event Selection} stage}

After the reconstruction stage, event are selected by applying the following requirements:
\begin{enumerate}
\item At least one jet with $P_T> 140$ GeV.
\item $E^{miss}_T > 140$ GeV, the {\it high } $E^{miss}_T$ region, to discriminate the signal from SM process.
\item The difference in azimuthal angle $(\Delta \phi)$ between the missing transverse momentum and each of up to four highest-$P_T$ jets with $P_T>50$ GeV is required to be larger than 1.0.
\item Candidates events are required to have no electron and no muon (lepton veto).
\end{enumerate}
At the generator level stage, the event selection follows the same criteria, except that the object definition is slightly different. The generator level missing transverse energy (dubbed ``Offline Missing Energy" by ATLAS) is defined as the the vector sum of the transverse momentum of neutrinos, neutralinos and charginos (the tracklet $p_T$ is not used). Generator level jets are defined using the  the anti-kt algorithm with a radius parameter of 0.4 over all particles except for muons, neutrinos, neutralinos and charginos with $c\tau$ above 10 mm.
Defining
\begin{itemize}
\item $N$ as the total number of chargino events,
\item $N_{gl}$ as the number of chargino events passing the Generator-Level kinematic selection,
\item $N_{es}$ as the number of chargino events passing the Event Selection,
\end{itemize}
the event acceptance $E_A$ and event efficiency $E_E$ are given by
\begin{equation}
E_A = \frac{N_{gl}}{N}  \qquad \text{,}\qquad  E_E = \frac{N_{es}}{N_{gl}}
\end{equation}
We stress that the model dependent quantities $E_E$ and $E_A$ can be computed directly from Monte Carlo simulation, as they do only involved standard reconstructed objects (no requirement on tracklets).

It is important to note that $E_E$ could be larger than one, because an event could be failing the generator-level cuts while passing the reconstructed level selection due to object resolutions. As a concrete example, a signal event could have a leading jet of $p_T = 135$ GeV at parton level and $p_T = 145$ GeV at reconstructed level. Such an event would not be included in $N_{gl}$, but would be part of $N_{es}$.

\subsubsection{The  {\it Tracklet Selection} stage}

After the Event selection stage, the ATLAS collaboration applies a series of requisites to the tracklets, in order to reduce the expected background. Due to the inherent nature of the tracklet as a reconstructed object, we can not reproduce this part of the analysis. Hence, we need to resort to the public information provided by ATLAS in the auxiliary material~\cite{1641262} in order to validate our analysis. The tracklet selection of ATLAS requests
\begin{itemize}
\item \textbf{Isolation and $P_T$ requirement}
     \begin{itemize}
	\item The separation $\Delta R$ between the candidate tracklet and any jet with $P_T>50$ GeV must
be greater than 0.4.
 	\item The candidate tracklet must have $P_T>20$.
	\item The candidate tracklet must be isolated. A track or tracklet is defined as
isolated when the sum of the transverse momenta of all standard ID tracks with $p_T > 1$ GeV and
$|z_0 \sin({\theta})| < 3.0$ mm in a cone of $\Delta R = 0.4$ around the track or tracklet, not including the $p_T$ of the candidate track or tracklet, divided by the track or tracklet $p_T$, is small: $p_T^{cone40}/p_T < 0.04$.
	\item The $P_T$ of the tracklet must be the highest among isolated tracks and tracklets in the event.
     \end{itemize}
\item \textbf{Geometrical acceptance}
     \begin{itemize}
	\item The tracklet must satisfy $0.1 < |\eta|<1.9$
     \end{itemize}
\item \textbf{Quality requirements}
     \begin{itemize}
	\item The tracklet is required to have hits on all four pixel layers.
	\item The number of pixel holes, defined as missing hits on layers where at least one is expected given the detector geometry and conditions, must be zero.
	\item The number of low-quality hits associated with the tracklet must be zero.
	\item Tracklets must satisfy requirements on the significance of the transverse impact parameter, $d_0$ , $|d_0|/\sigma(d_0) < 2$ (where $\sigma(d_0)$ is the uncertainty in the $d_0$ measurement), and $|z_0 \sin(\theta)| < 0.5$ mm. The $\chi^2$-probability of the fit is required to be larger than 10\%.
     \end{itemize}
\item \textbf{Disappearance condition}
     \begin{itemize}
	\item The number of SCT hits associated with the tracklet must be zero.
     \end{itemize}
\end{itemize}
This selection contains criteria that are impossible to employ in an independent analysis. Thus, in contrast, the simple Generator-Level selection is defined as follows
\begin{itemize}
\item $P_T > 20$ GeV.
\item $0.1 < |\eta| < 1.9$.
\item $122.5 \text{ mm} < R < 295 \text{ mm}$, where $R$ is the decay position defined as the cilindrical radius relative to the origin.
\item $\Delta R >0.4$ between the chargino and each of the up to four highest-$P_T$ jets with $P_T>50$ GeV.
\end{itemize}
In total analogy with the ``Event selection'' stage, we will introduce the following quantities:
\begin{itemize}
\item $n_{gl}$ as the number of charginos which pass the Generator-Level tracklet selection in events which pass the Event Selection,
\item $n_{rec}$ as the number of reconstructed events where at least 1 chargino is identified,
\item $n$ total number of charginos in events which pass the event selection.
\end{itemize}
From these, the tracklet acceptance $T_A$ and tracklet efficiency $T_E$ are computed as
\begin{equation}
T_A = \frac{n_{gl}}{n}  \qquad \text{,}\qquad  T_E = \frac{n_{rec}}{n_{gl}}
\end{equation}
In order to calculate $n_{rec}$, we need to use the $T_A T_E$ efficiency heatmap provided by ATLAS in the auxiliary material\cite{1641262}, where a given $\eta$ and radial decay distance $r$ the product $T_A T_E$ is provided.
As a final ingredient, the ATLAS collaboration provides the tracklet $p_T$ efficiency $P$, which is the probability that a tracklet passing the acceptance condition will have $p_T > 100$ GeV~\footnote{In the ATLAS study, this quantity is indistinctively called $P$ and $T_P$. We thank Ryu Sawada for clarifying the confusion.}.

With all the ingredients at hand, we can then explicitly write down the probability for a parton level event with $N$ charginos to have at least one reconstructed tracklet, which is given by
\be
1 - p(N, 0) = E_E E_A (1 - (1-T_A T_E P )^N )
\ee
where $p(i,j)$ is the probability that a parton level event with $i$ charginos yields $j$ reconstructed tracklets in the final state. This results coincides exactly with that quoted by ATLAS in footnote 5 of their paper, and thus provides an additional sanity check to our understanding of the analysis description.

In order to compute $n_{rec}$ from our parton level events, we proceed as follows. We consider that in the $i$th event, we have at most two charginos, which have a value of $T_A T_E$ given by the ATLAS heatmap which for short we call $\epsilon_1$ and $\epsilon_2$. Hence for this event we have
\begin{equation}
p(1,1) = \epsilon_1, \qquad p(2,1) = \epsilon_1 (1- \epsilon_2) + \epsilon_2 (1- \epsilon_1), \qquad p(2,2) = \epsilon_1 \epsilon_2 \, ,
\end{equation}
where to keep a simple notation we will omit an event dependent subscript "i" on each $p(a,b)$ function. Summing over all events we have that $n_{rec} = \sum \left( p(1,1) + p(2,1) + p(2,2) \right)$, allowing us to compute, for a given point in the ($m_{\chi}$-$c\tau$) plane, $T_A$ and $T_E$.

 \subsection{Validation in the AMSB scenario. }

 The ATLAS study uses as benchmark the minimal Anomaly Mediated Supersymmetry Breaking (AMSB) scenario~\cite{Giudice:1998xp,Randall:1998uk} where $\tan{\beta} = 5$, the universal scalar mass is set to $m_0 = 5$ TeV, and the sign of the higgsino mass term set to be positive. We performed a scan of chargino masses between 91 GeV and 700 GeV and lifetimes between $10^{-2}$ and 10 ns. Our signal simulation uses up to one additional parton in the matrix element with CalcHEP 3.7.5~\cite{Belyaev:2012qa}, using the AMSB implementation (http://hepmdb.soton.ac.uk/hepmdb:1013.0145) for parton level events, using PYTHIA v8.2.44~\cite{Sjostrand:2014zea} for parton shower and hadronization, and finally using Delphes 3.4.1~\cite{deFavereau:2013fsa} to simulate the detector effects, employing the default ATLAS card. We employ the {\tt NNPDF23$\_$lo$\_$as$\_$0130$\_$qed} parton distribution functions~\cite{Ball:2014uwa}, and a QCD scale equal to the invariant mass of the pair of winos was used for the calculation of the cross section at leading order (LO) and corrections at next-to-leading order (NLO) in the strong coupling constant were obtained using Prospino2~\cite{Beenakker:1996ed}.

The collaboration has chosen three benchmark points to showcase the efficiencies and acceptances discussed in the previous subsection, making public the SLHA cards for each of these points. We have found this to be a good practice and highly useful to validate our event generation pipeline, and we believe that in the spirit of the recommendations from the reinterpretation forum~\cite{Abdallah:2020pec}, signal cards should be made public whenever possible. There are two important effects on the signal samples that we would like to discuss in more detail in the next paragraph, namely the impact of smearing the parton-level chargino track, and the effect of combining the signal fixed-order calulation with the parton shower event generator.

\begin{figure}[htb]  
\centering
{\includegraphics[width=0.73\textwidth]{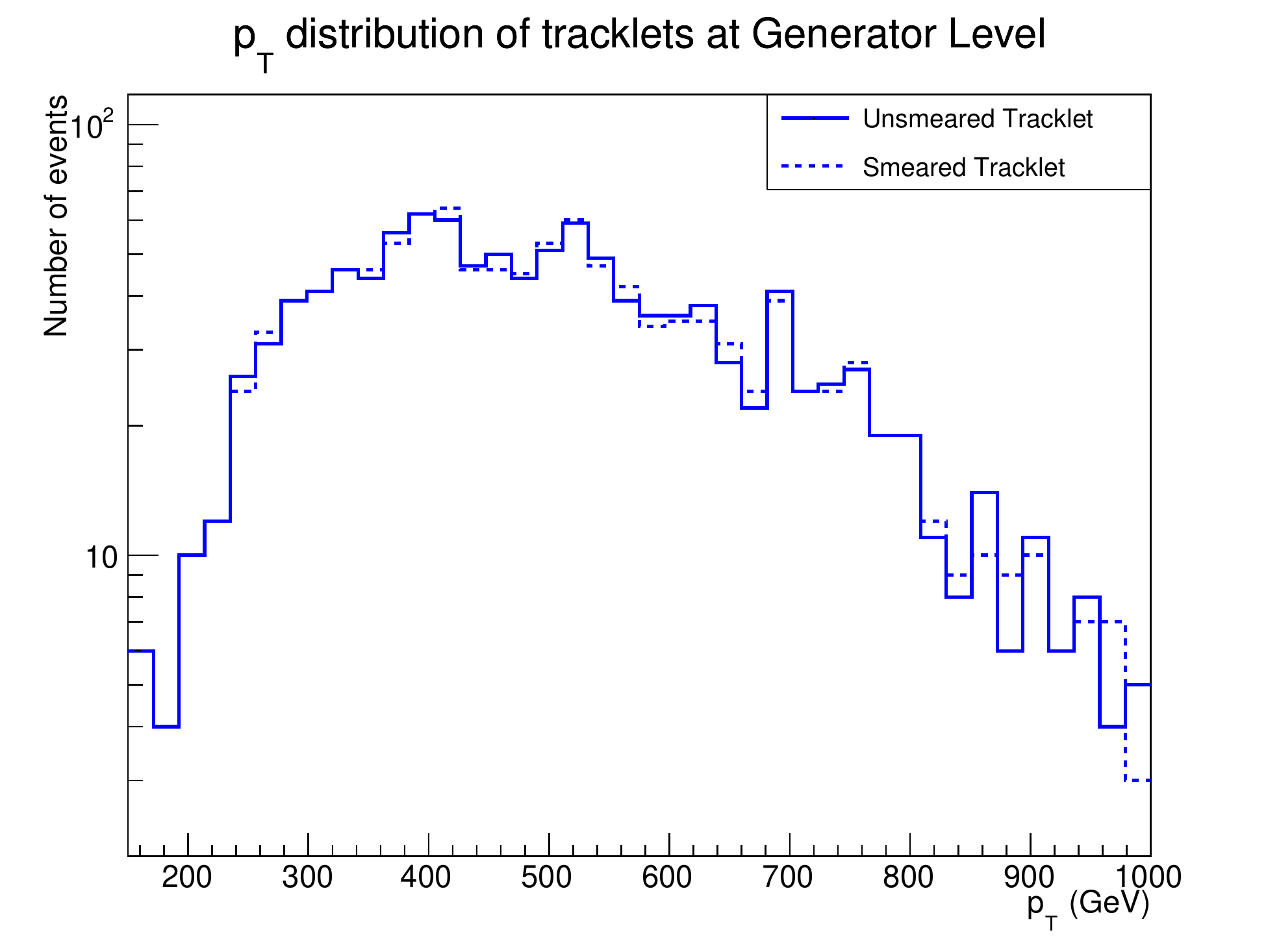}}%
\caption{Transverse momenta distribution $p_T$ of the chargino at the truth level (solid line) and smeared (dashed line) using a random scaling with 15 \% amplitude.} \label{fig:smeared}
\end{figure}

 We start with the effects of smearing.  We have checked that {the energy smearing, implemented by} multiplying each chargino four momenta by a factor of $1 + \Delta r / \sqrt{E}$ with $\Delta = 0.15$~\footnote{As a comparison, the smearing of a charged hadron with $p_T = 100$ GeV in the Delphes ATLAS card is 13 (17) \% for $|\eta| \in [0, 0.5] ( [0.5, 1.5])$.} and $r$ a random number flatly distributed between 0 and 1, does not have a {visible} impact in the $p_T$ distribution of the chargino, as displayed in figure~\ref{fig:smeared}. We have explicitly checked that the generator level instance of the tracklet selection does not change when considering the smeared sample and the original one, differing in less than 0.1~\%. We thus conclude that for disappearing track studies and electroweak pair production, smearing the disappearing track $p_T$ is not necessary.

Since both event and tracklet selection put strong constraints on jets, as well as include non-trivial cuts on correlations between jets and electroweak objects, multi-jet activity in the signal process has to be carefully modelled. Thus, multi-jet merged calculations are mandatory to obtain a tree-level accurate description of radiative spectra, minimizing the impact of (parton shower) approximations on the signal description. Multijet merged calculations are developed and tested for SM background processes. The typical renormalization- and factorization scales for these processes are of the order of the electroweak scale, so that the phase space for additional QCD radiation tends to be moderately small. For example, the ``natural'' scale of vector-boson + jets backgrounds is typically of the order of the vector boson mass, such that the impact of high parton multiplicity configurations on relatively inclusive observables (such as the boson rapidity) is moderate. This leads to relatively robust predictions for inclusive observables, and good agreement between different merging schemes. The differences between models appears at higher jet multiplicity, but the overall effect of these multiplicities is limited by the moderate value of the ``natural'' scale. The signal process at hand exhibits a large scale for QCD emission, such that the impact of higher-multiplicity calculations is not necessarily small. Thus, a robust prediction of the signal is not guaranteed~\cite{Dreiner:2012sh}, and an estimate of the uncertainty due to matching becomes mandatory in this case. We use two different multi-jet merging schemes to estimate the size of this uncertainty: MLM jet matching~\cite{Mangano:2001xp}, and CKKW-L multijet merging~\cite{Lonnblad:2001iq,Lonnblad:2011xx}.\footnote{This study lead to the identification of several critical errors in both the MLM and CKKW-L implementation in Pythia, $a)$ regarding the definition of processes with BSM resonance chains in the CKKW-L scheme, and $b)$ in the event rejection procedure, which is necessary to produce no-emission probabilities, for both merging schemes. The necessary corrections will be included in an upcoming Pythia release (tentative version 8.245).} As shown in detail in Appendix~\ref{app:matching}, 
distributions for leading jet transverse momentum and for 
transverse momentum of $\chi^+\chi^-$  pair indeed differ between two merging schemes. 
Differences arise because of several aspects of the merging of higher multiplicities: the mechanism to treat events that do not allow the interpretation as produced by a sequence of QCD emissions with decreasing hardness, the scheme how to set factorization scales for multi-jet events, and the procedure how to assign dynamic renormalization scales. In particular, factorization scales are not set on the basis of jet clustering in MLM, while the renormalization scales are set using the nodal jet separation values when clustering the partons into jets in the kt-algorithm. The latter procedure can potentially result in small renormalizaton scales for events that do not allow an interpretation as ordered sequence of QCD emissions, resulting in artificially large running-coupling values. Similar effects have been investigated (and corrected) in the CKKW-L scheme~\cite{Christiansen:2015jpa}. The shape of the MLM prediction could be adjusted a posteriori to some degree by ``tuning" the matching scale.
The CKKW-L scheme yields a smooth, physical $p_T$ distribution irrespective of the merging scale, while the event rejection in MLM jet matching induces visible matching artifacts in the transition region when the merging scale is not ``tuned'' so that the prediction recovers a target baseline. 
In case of pair $D^+D^-$ production, the merging scale value has potentially large uncertainty, since values that might be considered reasonable range from the transverse momentum of the $D^+D^-$ pair (of the order of 100 GeV) to the invariant mass of $D^{+}D^{-}$ pair (of the order of TeV). Therefore, for a proper handling of the transition region, we have adopted the CKKW-L scheme throughout the whole article. We use MLM scheme as a cross-check to assess if our conclusions depend heavily on the (highly scheme-dependent) dynamics of regions with moderate jet separation.

We reproduce the information on the acceptances and efficiencies in Table~\ref{tab:accep}, together with our own results which are displayed in parenthesis. We also present the ratio of the product $E_A E_E T_A T_E$ between ATLAS and our simulation. We see that we err by up to 20 \%, which is acceptable for a simplistic parton level simulation of the signal. We also see that our rate is lower than the corresponding one from ATLAS, and hence in these particular points our simulation gives a conservative estimate of the ATLAS result.

 \begin{table}[ht]
\label{Event_Tracklet}
\begin{center}
  \begin{tabular}{c|c|c|c|c|c|c|}
			\hline
			\hline
 \multicolumn{2}{c|}{Signal} & \multicolumn{2}{c|}{ Event} & \multicolumn{3}{c|}{ Tracklet}  \\ \hline
  $m_{\chi}^{\pm}$ (GeV)               &  $\tau$ (ns) &  $E_A$ &  $E_E$ & $T_A$ & $T_E$ & $\Delta $\\  \hline
 $400$ &     0.2     &   0.09  (0.09)  &  1.03 (1.03)  &  0.07 (0.08)  &  0.47 (0.44) &  1.211  \\
 $600$ &     0.2     &   0.12  (0.10)  &  1.05 (1.03)  &  0.05 (0.06)  &  0.48 (0.44) &  1.289  \\
 $600$ &     1.0     &   0.11  (0.10)  &  1.03 (1.03)  &  0.20 (0.22)  &  0.47 (0.43) &  1.169  \\ \hline
  \end{tabular}
 \end{center}
\caption{Event and tracklet  acceptances and efficiencies (see main text for definitions) 
for  some signal models, as an example.
Our results are shown within the parenthesis. The final column $\Delta$ show the ratio between the ATLAS values and our own. The overall error is around 20 \%, which is acceptable for a simplistic parton level simulation of the signal. We also note that the bulk of the difference originates from the ``tracklet" stage.}
	\label{tab:accep}
\end{table}

\begin{figure}[htb]
\centering
{\includegraphics[width=0.53\textwidth]{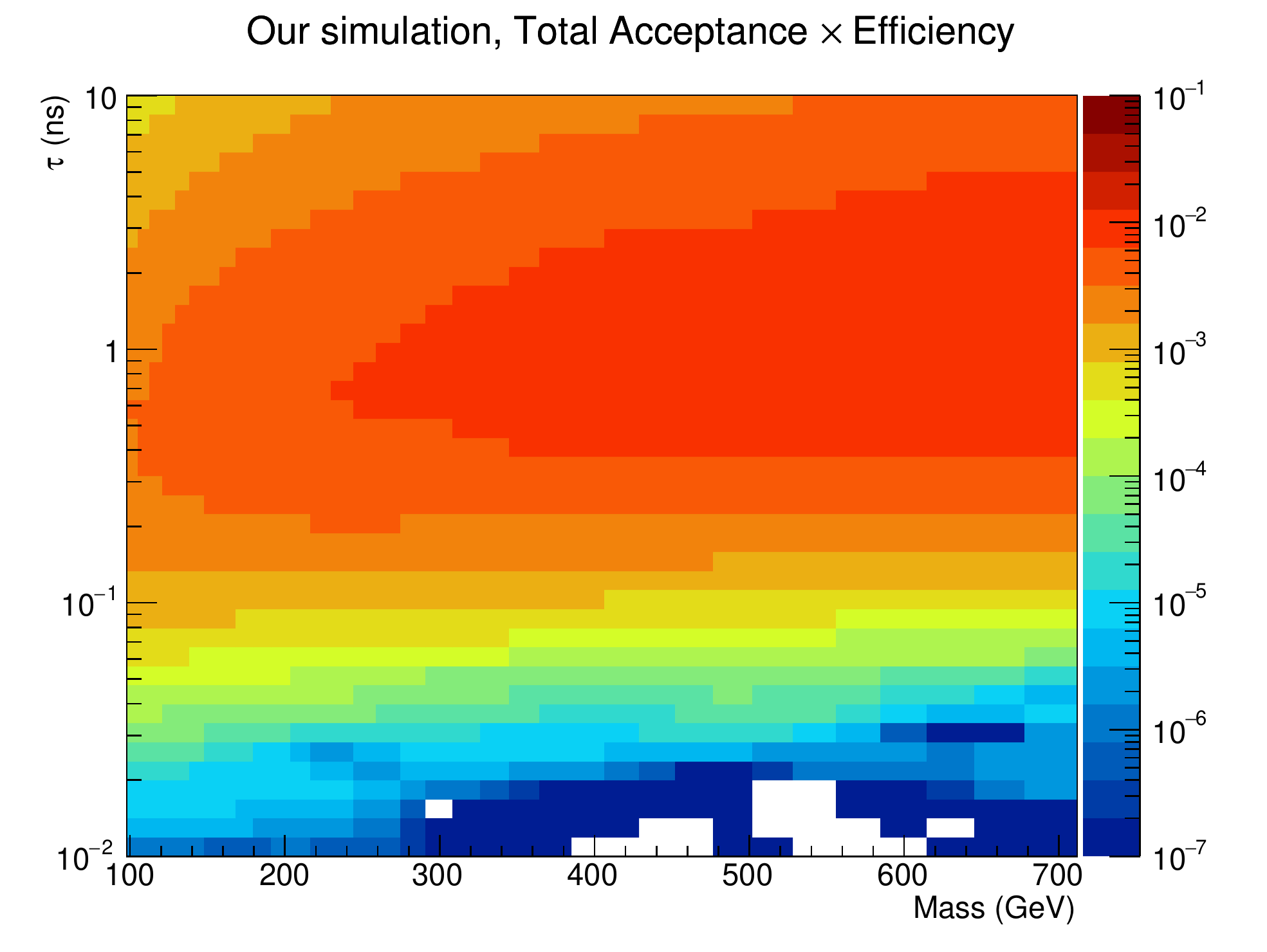}}%
{\includegraphics[width=0.53\textwidth]{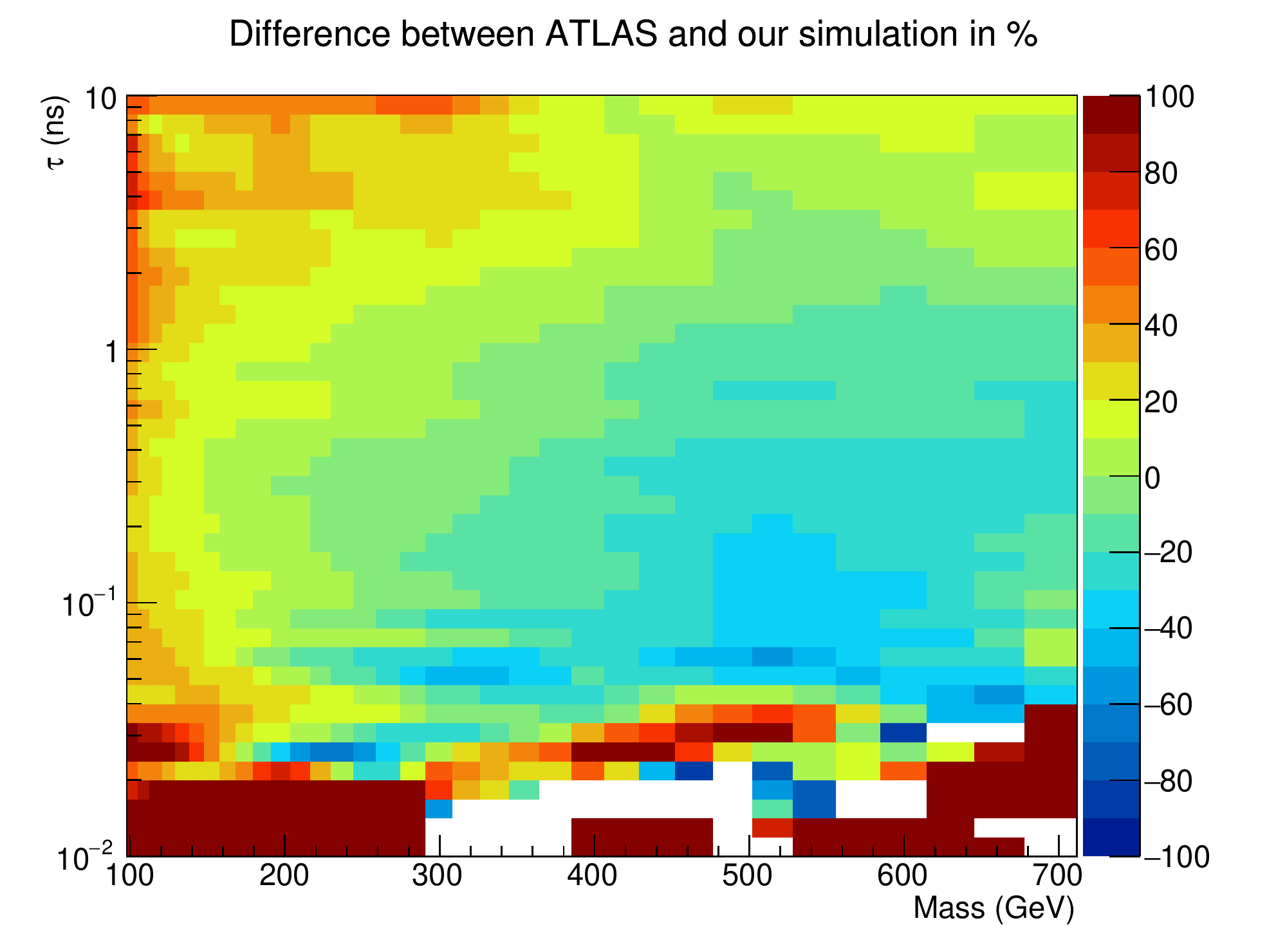}}%
\vskip -0.5cm\hspace*{-3cm}(a)\hspace*{0.48\textwidth}(b)
\caption{(a) Total acceptance $\times$ efficiency in the electroweak channel ($E_A \times E_E \times T_A \times T_E $) from our simulation. b) Difference of the Total Acceptance $\times$ efficiency in \% between ATLAS  and our simulation.} \label{fig:tate}
\end{figure}

Since the ATLAS collaboration provided 2-D binned results for the product $T_A T_E$ in the   ($m_{\chi_1^{\pm}} - c\tau$) plane we thus show our own results, and the ratio between those and the reported ATLAS values in figure~\ref{fig:tate}. We this confirm that in most of the parameter space we are within a 20 \% error on the efficiency, while these values degrade when going to the edges of the scanned space.

In order to obtain exclusion limits that we can compare with the published ATLAS results, we still need to discuss the background events. The dominant backgrounds for the disappearing track signature are mostly of instrumental nature, and hence cannot be simulated with an event generator, but are rather obtained from the experimental data itself. In fact, the leading background after the tracklet selection is given by fake tracklets, namely, those coming from random hits of particles in the pixel layers. We need then to rely on the published $p_T$ distribution of the background done by the ATLAS collaboration, which indicates a total observed (expected) number of background events of 9 (11.8) for $p_T > 100$ GeV.

Finally, we need to further apply $P = 0.57$ to every chargino in our events.\footnote{We note that this value has only been presented for the three benchmark points. We assume it to be flat throughout the whole parameter space. } We note that the ATLAS collaboration has not taken full advantage of the events featuring two disappearing tracks. In such a case they have decided to keep only the hardest tracklet in the event, hence applying a probability of $p(1,1)$ to it.

\begin{figure}[htb]
\centering
\includegraphics[width=0.70\textwidth]{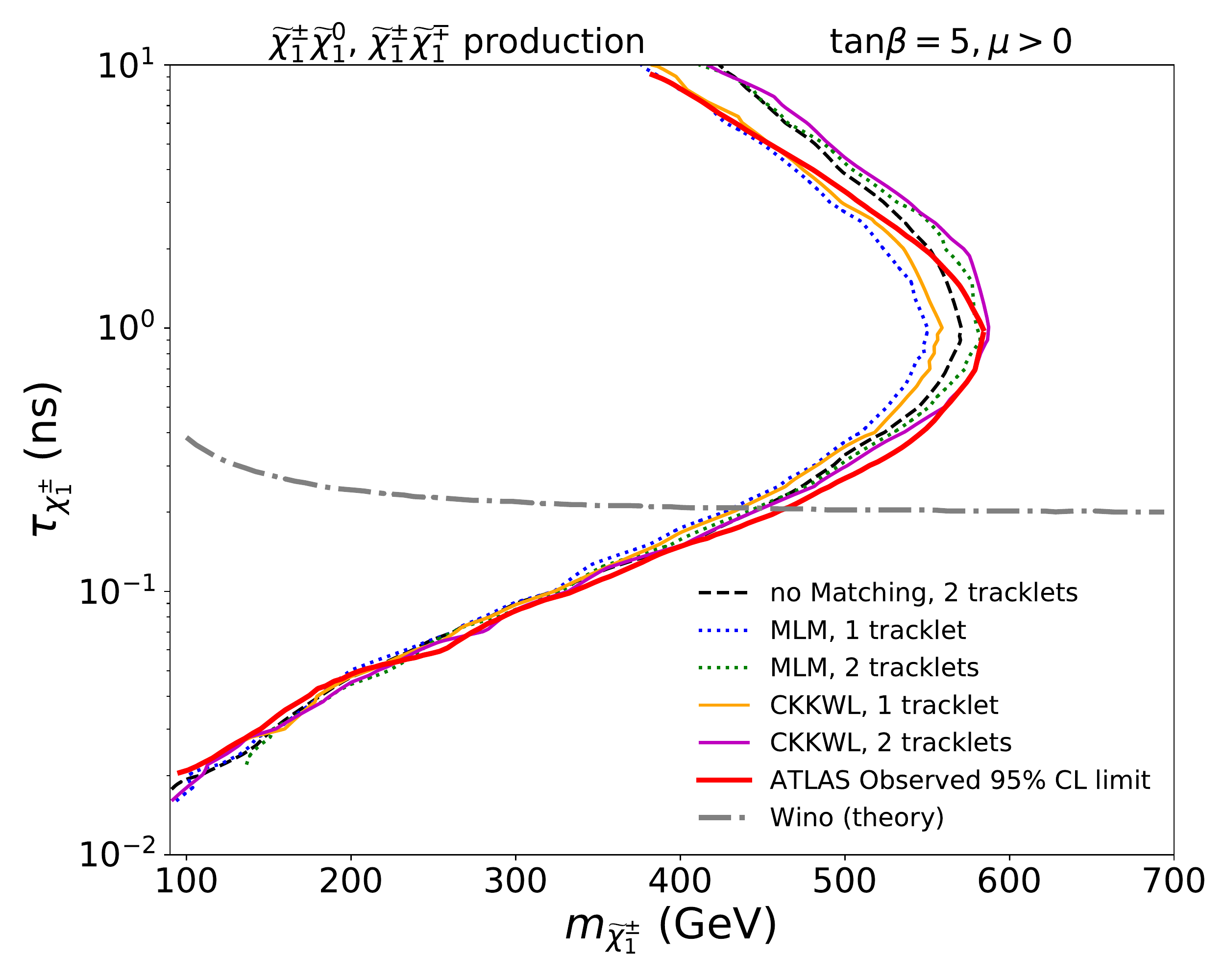}%
\caption{Exclusion limits from ATLAS (red line) vs. our simulation considering five different procedures. The dashed black line is the limit without using matching or merging, the dotted blue line consider MLM matching and the probability of reconstruct just 1 tracklet per event, the dotted green line also consider MLM matching and the probability of reconstruct up to 2 tracklets per event, the continuous orange line consider CKKW-L merging and the probability of reconstruct just 1 tracklet and the continuous purple line consider CKKW-L merging and reconstructed up to 2 tracklets per event. Finally the grey dashed line showes the theory curve of the chargino lifetime in the almost pure wino LSP scenario at the two-loop level \cite{Ibe:2012sx}.}  \label{fig:Exclusion_limit}
\end{figure}

The final results of our validation are then shown in figure~\ref{fig:Exclusion_limit}, where we compared the results from our simulation with the ATLAS exclusion limit, which is shown in solid red. The black dashed line shows the results when not including extra radiation at tree-level accuracy through matching or merging, i. e. all radiation is modelled  by parton showering alone. For both matching/merging procedures, we show two results, which differ if more than one tracklet is reconstructed: the result of using the hardest tracklet (as suggested by ATLAS), and the result when using both tracklets.

We immediately note that for lower lifetimes, the curves do not differ strongly from each other. However, for $\tau \sim $ ns, we can note large differences. We checked that if we use the probability to detect up to 2 tracklets in each event in samples with up to 1 parton, we obtain similar results as if we had used only the hardest tracklets in the event in samples with up to 2 partons, as ATLAS did in their analysis. With this we conclude that generating samples with up to 1 extra parton in the matrix element is enough to reproduce ATLAS limits, but we encourage the use two tracklet events which obviously would yield a stronger bound.

Using the CKKW-L merging scheme and up to two tracklets yields the more accurate agreement with the ATLAS result. The MLM scheme with 2 tracklets performs slightly worse, while only using one tracklet gives a lower exclusion limit, being overconservative. In particular, for a nominal wino lifetime of 0.2 ns we obtain an upper bound on the Wino mass of 444 GeV, instead of the 458 GeV result from ATLAS, hence a 3\% difference in mass and 15\% in the signal cross section we are sensitive.
We have also verified that the use of an appropriate matching procedure is necessary to obtain consistent exclusion limits, especially for wino masses above 450 GeV.

\begin{figure}[htb]
\centering
\includegraphics[width=0.70\textwidth]{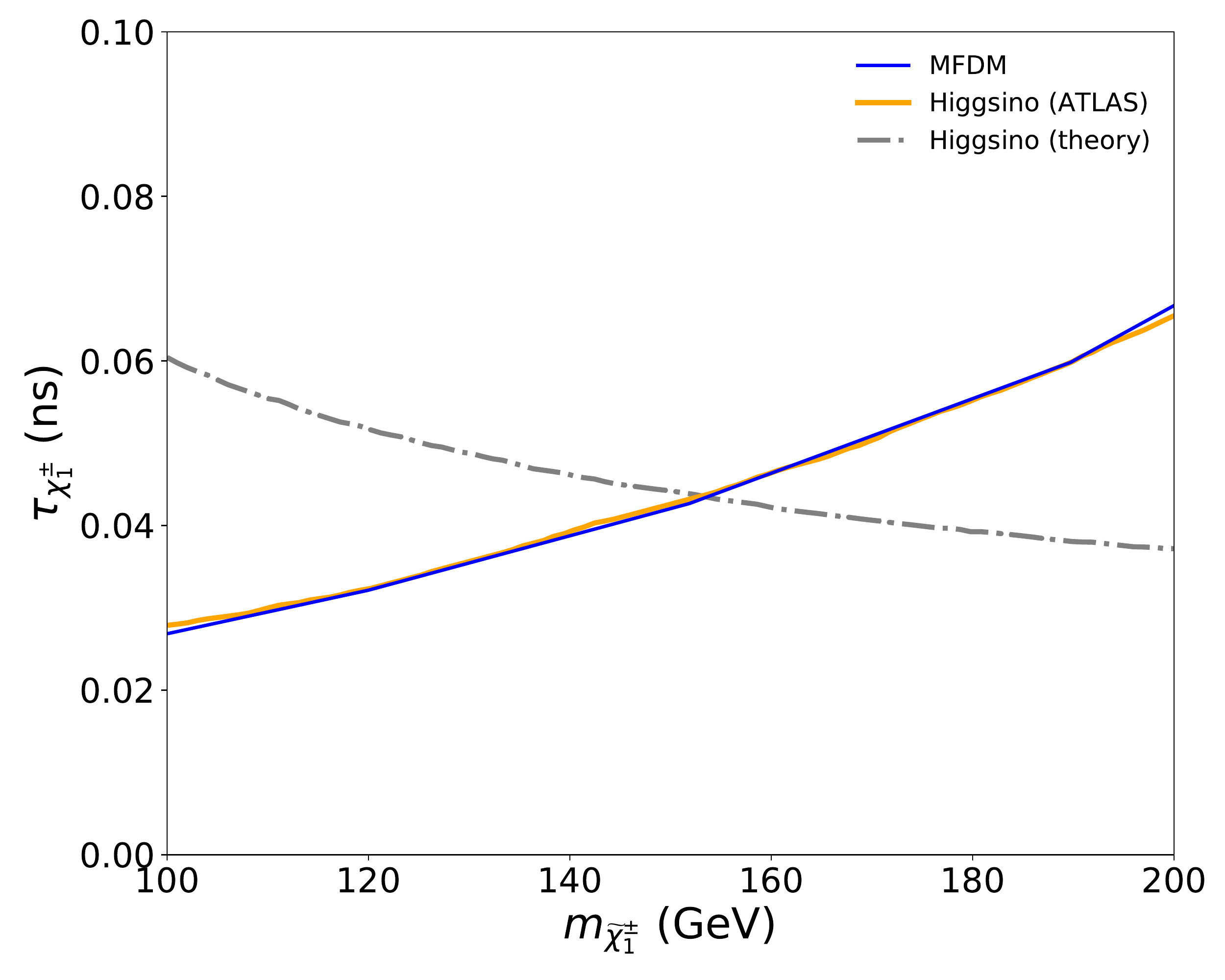}%
\caption{Comparison between the official ATLAS reinterpretation of the disappearing track study in the Higgsino scenario (solid orange) and our recasting procedure. The solid blue shows our results, using NLO cross-sections from PROSPINO.}
\label{fig:Higgsino_recast}
\end{figure}

The ATLAS collaboration has also interpreted the results of their study in the context of Higgsino dark matter~\cite{ATL-PHYS-PUB-2017-019}. Hence their results provide an additional check for our procedure. They have only displayed their recasting in the 100-200 GeV range, this is why we will only compare the Higgsino model in this range. We show our results and those of the ATLAS collaboration in figure~\ref{fig:Higgsino_recast}.

We see that there is an excellent agreement between our reinterpretation and the ATLAS results in the whole mass range, where the largest differences do not exceed 5\%. 
We see that for a fixed $c \tau$, ATLAS excludes a larger mass, hence our recasting turns up to be on the conservative side. 

We can close this section by concluding that our recasting procedure reproduces well the published ATLAS results for the wino and higgsino models. The Python code implementing this procedure, together with the corresponding instructions, is publicly available in~\cite{LLPrepo}. In the next section we will then apply this procedure to other models of dark matter.


\section{Reinterpretation for Minimal Models}
\label{sec:results}
In this section, we apply our validated recasting to a few selected examples of minimal models. We note that in many cases, this provides the best probe of the parameter space and the first \emph{direct} constraints on long-lived electrically charged particles.

In all our models, we have two kind of constraints, that arise from \emph{direct} searches and from \emph{indirect} effects. 
The current direct collider constraints tend to be mild and come mostly from LEP searches~\cite{Heister:2002mn,Abbiendi:2002vz,Abbiendi:2003yd,Abdallah:2003xe}. At the LHC, the production of dark sector particles would lead  invariably to $\met$ signals. We are focusing on cases where $\Delta M$ is small and hence $D^{\pm}$ is long-lived. Thus, any study in the $X + \met$ class (where  $X$ is some combination of SM particles) does not apply, \emph{if $X$ arises from intra dark-sector decays}, since $X$ will be too soft to be detected. Nonetheless, $X$ can arise through initial state radiation.
The highest rates for $X + \met$ signature is for the monojet final state (which is to a good approximation  independent of the small mass splittings in  the dark sector and does not depend on $c \tau$) which provides robust  lower limit on the dark sector mass scale  for a given model.~\footnote{For $c \tau \gtrsim 1 m$, there are important constraints coming from Heavy Stable Charged Particle Searches (HSCP)~\cite{CMS-PAS-EXO-16-036,Aaboud:2018hdl}. We note, however, that the focus of this paper is on tracks with nominal lifetime between 0.01 and 1 ns, namely a proper displacement of 3 mm - 30 cm. For a lifetime of 30 cm and electrically charged particles, the HSCP does not yield competitive constraints. }.

The indirect constraints that apply arise from either electroweak precision data (e.g: Peskin-Takeuchi $S,T, U$ parameters) or from 1-loop effects from our charged particles in $b \to s \gamma$ decays. We note that these constraints are weak, as they can be reduced by additional contributions not explicitly involved the dark matter dynamics. For instance, in supersymmetric models, this rare decay proceeds via a chargino-stop loop, and the stop sector does not play any role in the dark matter phenomenology. Due to their strong model-dependence, we will ignore these constraints in what follows.

\begin{figure}[htb]
\centering
\includegraphics[width=0.5\textwidth]{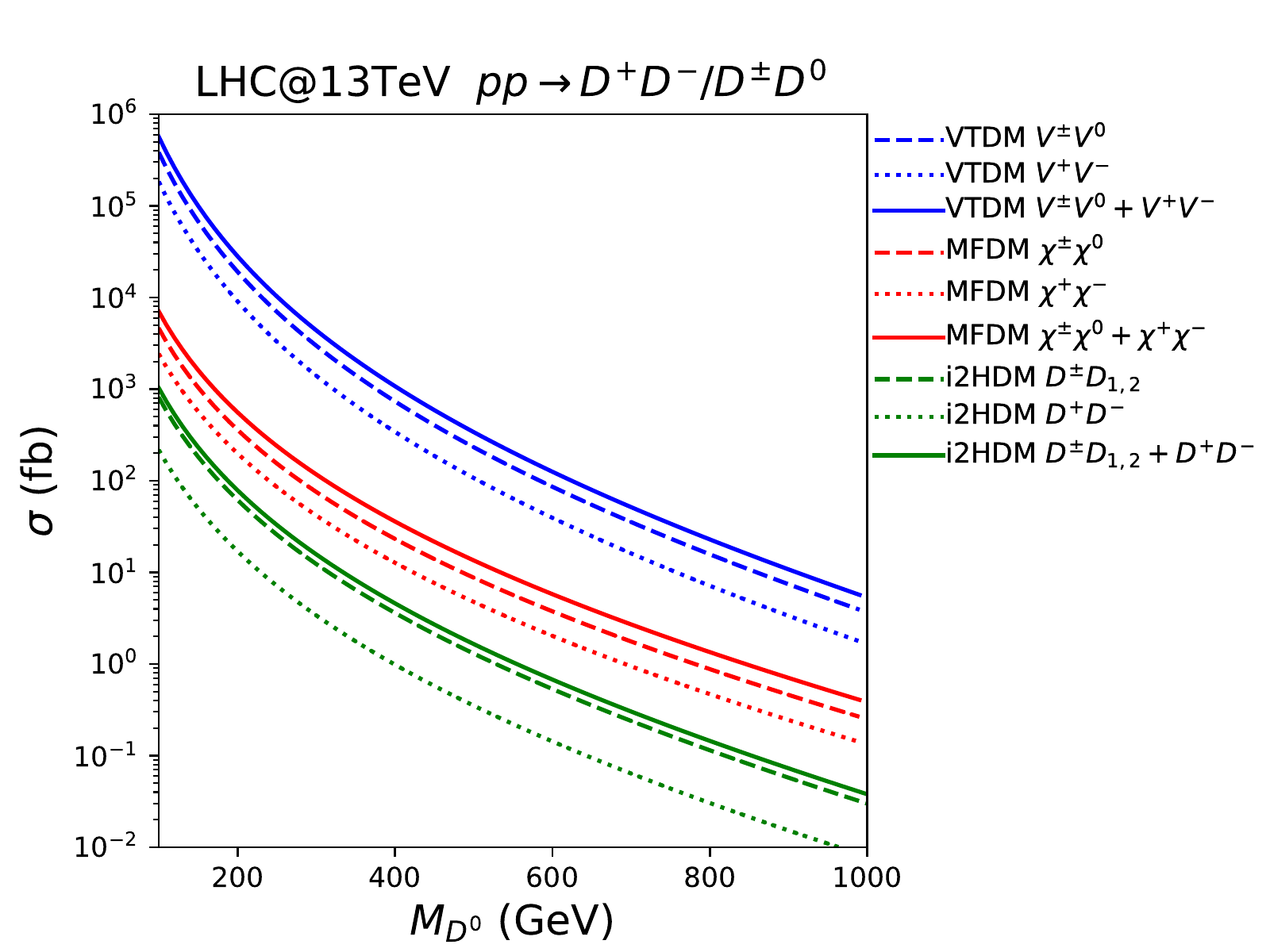}%
\includegraphics[width=0.5\textwidth]{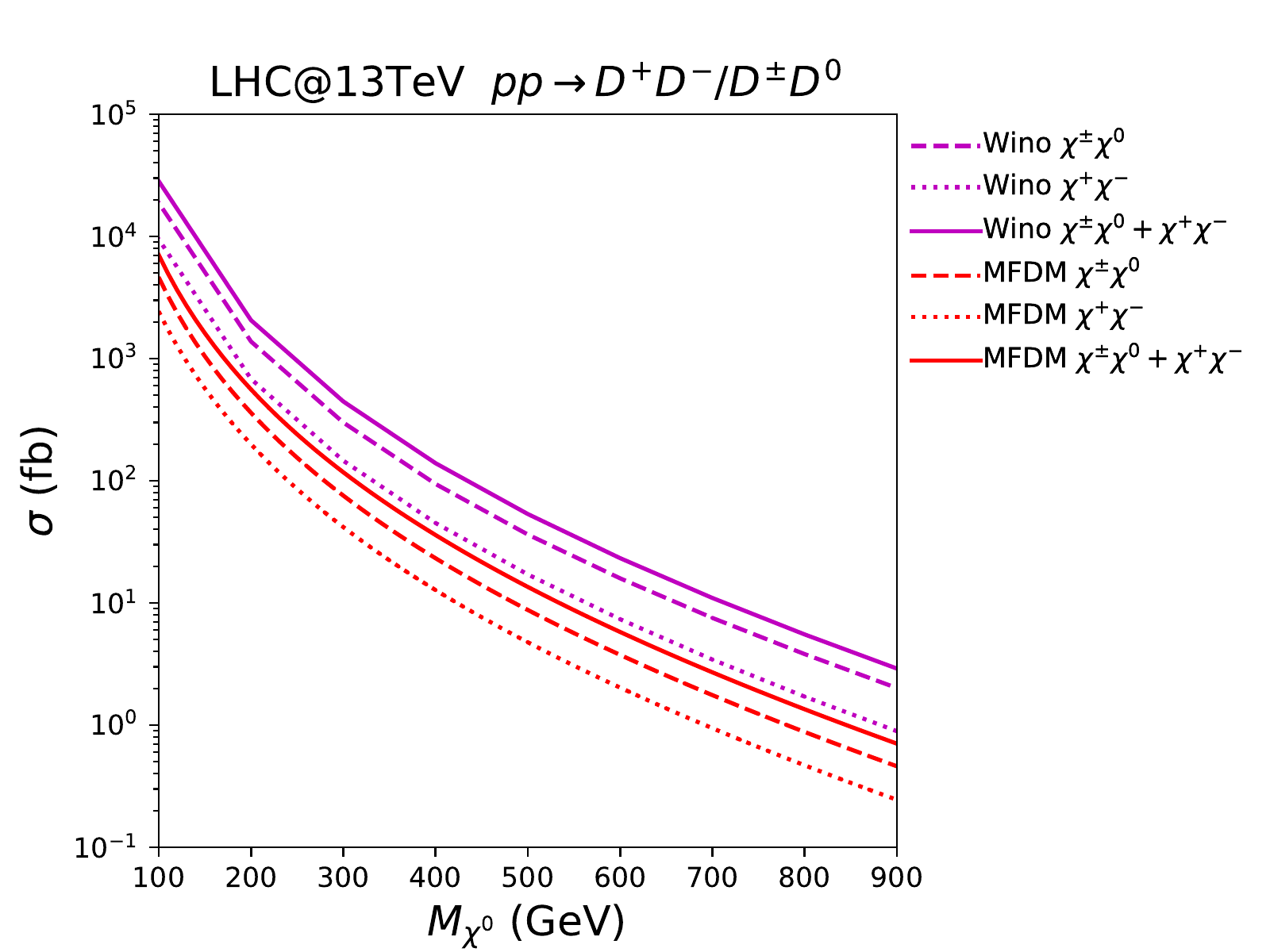}%
\vskip -0.2cm \hspace*{-5cm}(a)\hspace*{0.48\textwidth}(b)
\includegraphics[width=0.5\textwidth]{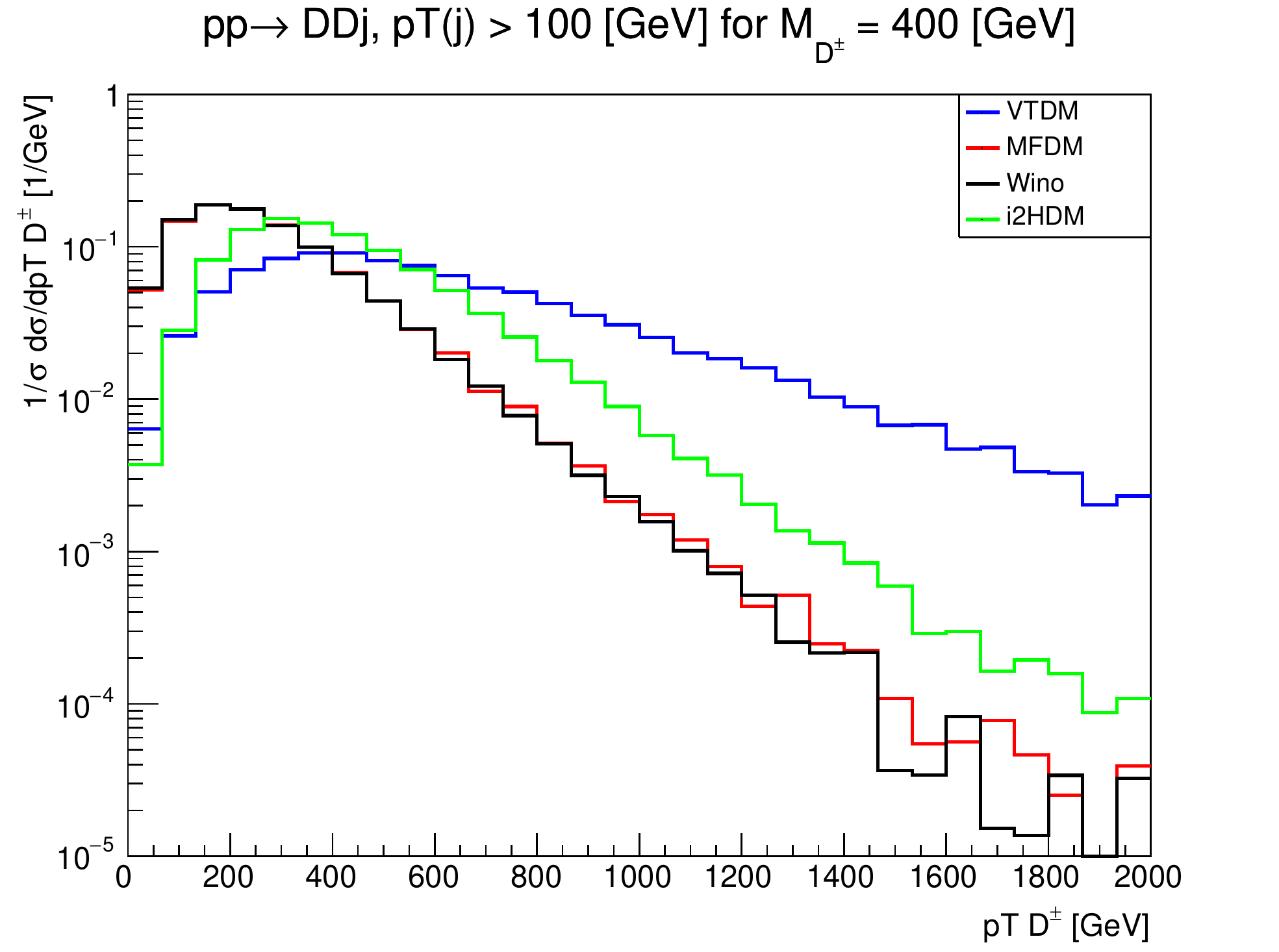}%
\includegraphics[width=0.5\textwidth]{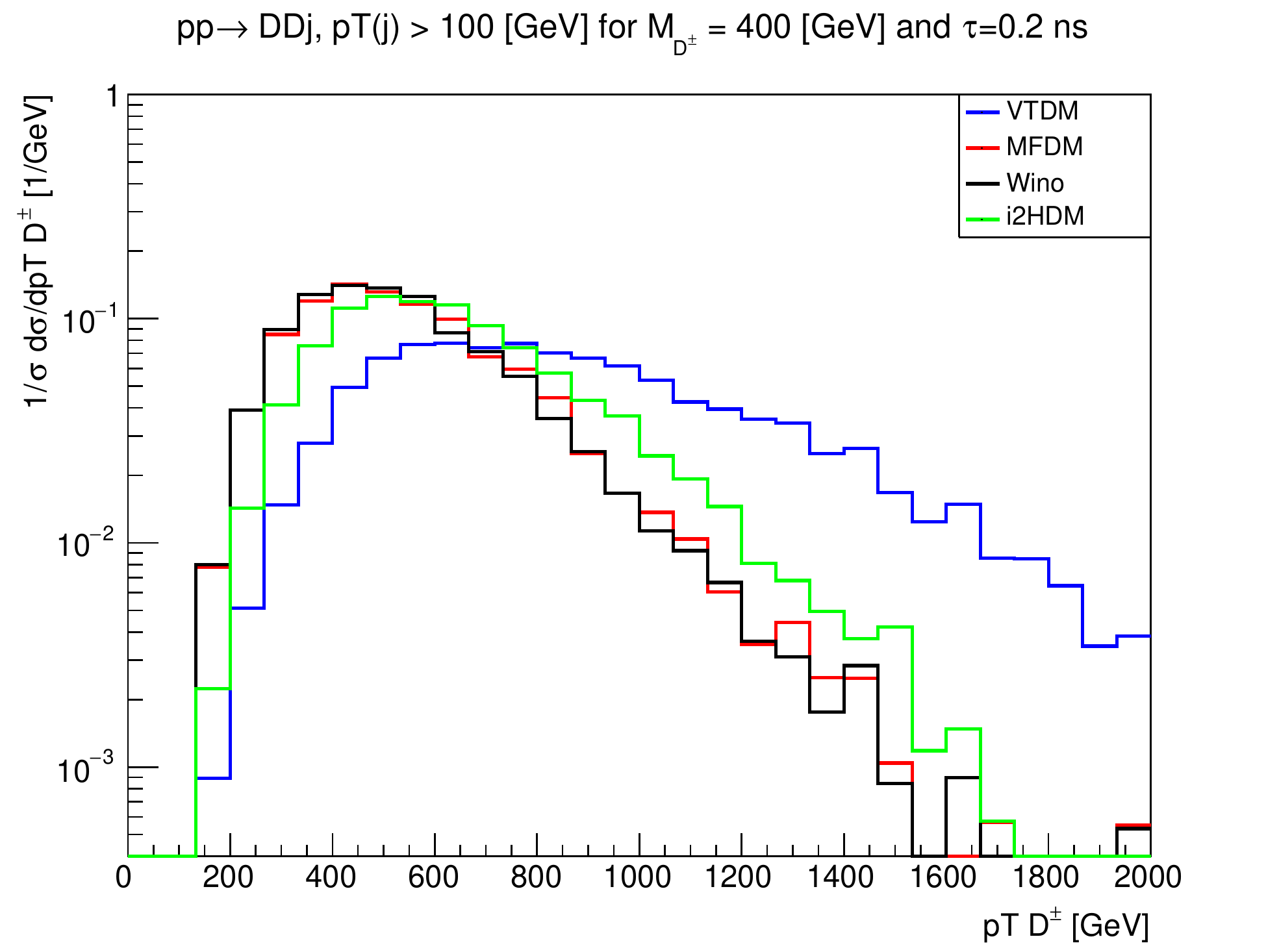}%
\vskip -0.2cm \hspace*{-5cm}(c)\hspace*{0.48\textwidth}(d)
\caption{(a) and (b) Production cross section for the charged-neutral ($D^{\pm}D$) and charged-charged ($D^+D^-$) pair production of dark sector particles in the models under consideration, as a function of the charged dark particle mass. (c) Transverse momentum distribution of the short-lived charged dark particle (``chargino") at the parton level. (d)  Transverse momentum distribution of the reconstructed charged track.
}
\label{fig:models_xs}
\end{figure}

We display in figure~(\ref{fig:models_xs}a) the production cross section of pairs of particles: (dotted line) charged-charged, (dashed line) neutral-charged and (continuous line) the sum of both contributions, for the vector (VTDM), fermion (MFDM) and scalar (i2HDM) model. In figure (\ref{fig:models_xs}b) we show the analogous plot, for the wino and MFDM models.

Furthermore, in the figure~(\ref{fig:models_xs}c) we show, for the $D D$ + jet process with $p_T(j) > 100$ GeV,  the $p_T$ distribution of the charged dark particle at parton level normalized to the cross section, while in the figure~(\ref{fig:models_xs}d) we convolute the experimental efficiencies obtaining the reconstructed charged track $p_T$ (note that the latter does depend on the lifetime of $D$, while the former does not). From the figure, we see that the spectrum is much harder for vectors than for scalars and fermions. Given that the vector model also enjoys the largest cross section, we can expect the most stringent exclusions to occur for the VTDM. The $p_T$ spectrum in the fermionic models is softer than in the i2HDM model. However, the cross section for fermionic scattering are larger than for fermion scattering (for i2HDM by an order of magnitude). Hence, we can naively expect that fermionic models will follow after VTDM in the hierarchy of constraints. Scalar models will presumably have the mildest constraints.

\begin{figure}[htbp]
\begin{center}
\includegraphics[width=0.70\textwidth]{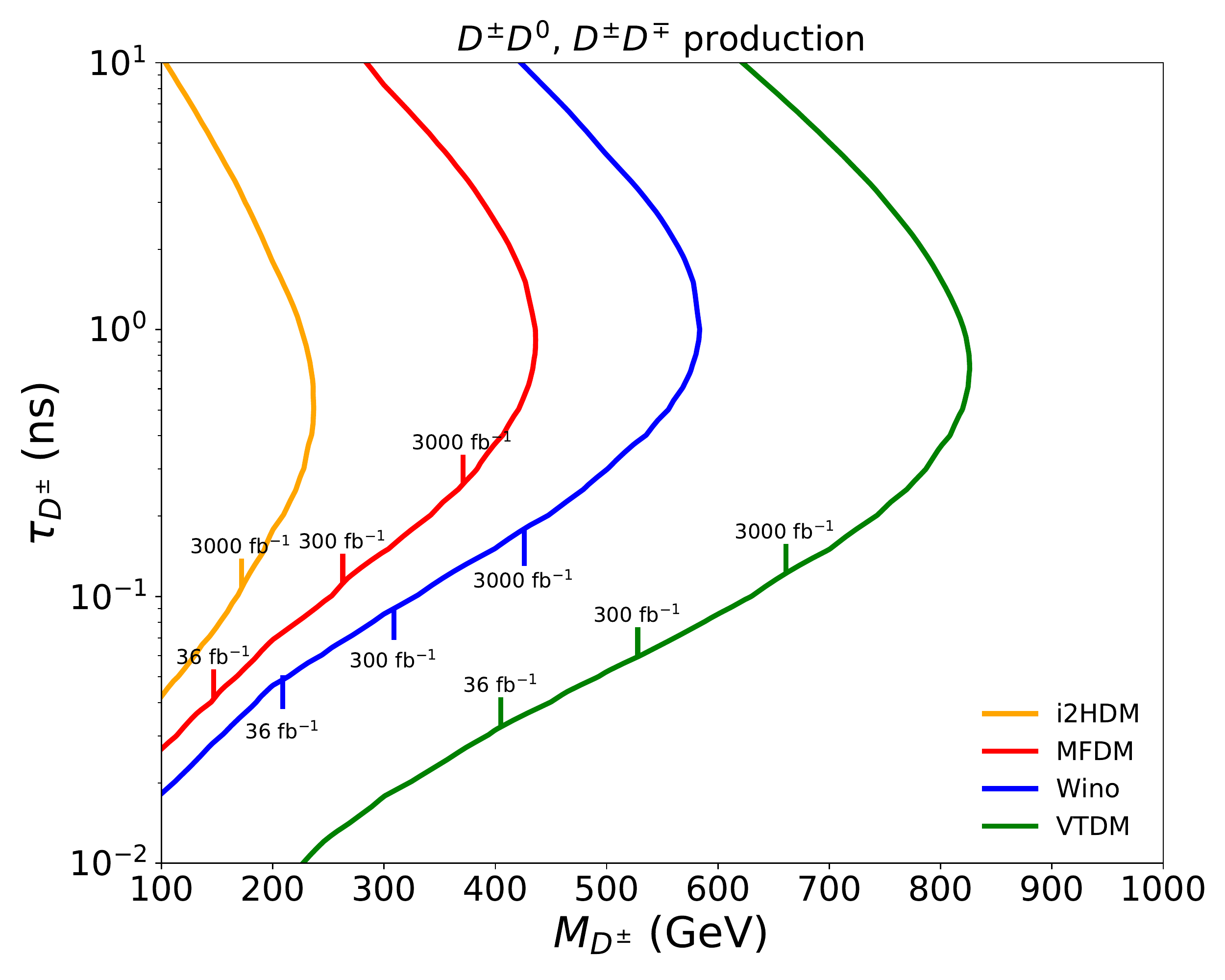}
\caption{Constraints on the parameter space of the dark sector models studies in this paper. The colored lines show the results of the reinterpretation of the disappearing track ATLAS study for the different models. The solid ticks indicate the constraints coming from LHC mono-jet searches at 36 fb$^{-1}$ and projections for 300 and 3000 fb$^{-1}$. } 
\label{fig:dt_exclusions_1}
\end{center}
\end{figure}

In figure~\ref{fig:dt_exclusions_1}, we present the current LHC potential to probe the  $(\tau_{D^\pm}-M_{D^\pm})$
	parameter space of the  MFDM, VTDM, wino and i2HDM models with the disappearing track signature. We further superimpose the limits from the current~\cite{Aaboud:2017phn} and future mono-jet searches as obtained using \cite{PhysRevD.99.015011} results.\footnote{We have verified for several benchnmark points that the results of \cite{PhysRevD.99.015011}are in good agreement with CheckMATE2~\cite{Dercks:2016npn}. }
The coloured lines show the bound obtained from our reinterpretation of the disappearing track search for each model.
The solid ticks indicate the corresponding limit from the LHC mono-jet searches for the specified luminosity.\footnote{
A comment on the reinterpretation of mono-jet searches for long-lived charged particles is in order, regarding how the $D^{\pm}$ particles pass the event selection, depending on their lifetime. Since missing energy is computed in~\cite{Aaboud:2017phn}  from visible calorimeter deposits, for very low lifetimes where $D^{\pm}$ gives only very little (or zero) pixel hits ($c \tau \lesssim 1$ mm) the prompt analysis can be directly applied. As the lifetime increases, however, the $D^{\pm}$ appears with more and more tracker hits, and even with calorimeter deposits due to the exponential decay tail. In that case, it is clear that the $D^{\pm}$ will not satisfy the loose jet selection criteria~\cite{ATLAS-CONF-2015-029} adopted in the mono-jet study, where such events are discarded. Hence, we expect that the mono-jet limit will degrade for larger lifetimes, however its proper assessment is outside the scope of this work.}.

We see that constraints derived from our reinterpretation of the disappearing track search  probe a vast region of the parameter space, well beyond other LHC or LEP searches. We stress again that for 36 fb$^{-1}$ LHC data  we have found  the limit  $m_\chi > 447$ GeV and $m_\chi > 152$ GeV  at 95 \% C.L for
fermionic triplet (MSSM Wino) and doublet (MSSM Higgsino) model respectively, which is in good agreement with the official results from ATLAS (460 GeV and 150 GeV respectively). Note that these current limits are even stronger that the projected HL-LHC mass reach. In the VTDM model, disappearing tracks set a lower bound on 530 GeV for the nominal lifetime of 0.06 ns ($c \tau \sim$ 2 cm ) \footnote{
The limit on DM mass from VTDM found in~\cite{Belyaev:2018xpf} was recently corrected (there was a typo in the analysis code) and  agrees well with the more accurate estimation we found in this paper.}
For the i2HDM model the strongest  limit is 237 GeV, which corresponds to a lifetime of 0.53 ns and $\Delta M = 0.157$ MeV. This limit is larger than the estimated HL-LHC reach of about 190 GeV using mono-jet\cite{PhysRevD.99.015011}.

 We summarize these results in table~\ref{tab:max_excluded}, where we present the maximum excluded mass for each model under our scrutiny.
\begin{table}[ht]
\caption{Maximum excluded mass for different DM models.} 
\begin{center}
  \begin{tabular}{c|c|c}
			\hline
			\hline
Models  & Mass (GeV) &  tau (ns)  \\ \hline\hline
i2HDM   &   237      &    0.5     \\ \hline
MFDM    &   436      &    0.9     \\ \hline
VTDM    &   822      &    0.7     \\ \hline 
WINO    &   587      &    1.0     \\  
  \end{tabular} 
	\end{center}
	\label{tab:max_excluded}
\end{table}
We see that our intuition regarding the constraint hierarchy for the different models is confirmed.

The high potential of this LHC LLP study shows the paramount importance of not only conducting this class of searches, but also of clarifying  the analysis assumptions (object definitions, model cross sections used) and the required ingredients (efficiencies) for its prompt reinterpretation in the context of arbitrary models.


\section{Conclusions}
\label{sec:conclu}
\label{sec:conclusions}
In this work we have performed detailed studies devoted to the reinterpretation of 
the current disappearing track searches at the LHC for a wide class of models, going beyond the vanilla examples of higgsino and wino dark matter.

We have validated our reinterpretation procedure by carefully following the ATLAS study~\cite{Aaboud:2017mpt}, and reproduced, to good accuracy, their published results for wino dark matter. By using the provided efficiency heatmaps in the mass-lifetime chargino plane, we were able to obtain a good agreement with the ATLAS results, being a few percent away in most of the parameter space, with the largest differences taking place for lifetimes of about 1 ns. We have further studied the impact of smearing the parton level charged track momenta, which we have found to be a sub-percent effect. An accurate description of multi-jet final states is necessary for a precise description of the data. We conclude that tree-level calculation for at least one additional hard jet is necessary to define the signal and that the CKKW-L merging scheme (which may be considered more predictive if additional QCD radiation can be very hard) allows to perform more accurate simulation of the  two-tracklet  system. The modelling of the transition region of moderate jet separation can be important, in particular before the tracklet selection, which led us to assess the matching-dependence of the results by using two different merging schemes. 
Finally, we have also validated our procedure for the higgsino study, following the ATLAS reinterpretation of their own results~\cite{ATL-PHYS-PUB-2017-019}.

With our validated procedure, we have reinterpreted the disappearing track search  in the context of several models of dark matter: the minimal Fermion doublet model (MFDM), the vector dark matter model (VTDM) and the inert 2-Higgs doublet model (i2HDM). Our results shown in figure~\ref{fig:dt_exclusions_1} are the core result of this paper. As byproducts of our analysis, we also provide a) our upper limits and efficiencies on the lifetime-mass plane of the different dark matter models considered here  and b) the python code used for our analysis, in~\cite{LLPrepo}, which is ready to run over event samples of arbitrary DM models.

It is worth stressing that while most of the information about the ATLAS study was publicly available, close contact with the experimentalists that carried out the study was still required, in order to understand a few crucial details. Our interaction with our experimental colleagues has been very fruitful, and helped considerably in explaining the study and in dispelling doubts. However, it would desirable that a reinterpretation of an experimental study does not require to consult with the experts from the experimental collaborations. Going into details, the multiple definitions of efficiencies and acceptances were not immediately clear to us. As these are terms that are burdened by various interpretations, we encourage the experimental collaborations to define these quantities with mathematical formulae that are universally understood and not prone to a language interpretation. These considerations also apply to the definition of observables at the different simulation and reconstruction levels. Furthermore, a key ingredient for the comparison with the benchmark model(s) chosen by the collaboration is to also report the assumed cross section values, as often it is not clear the exact parameters used as input to the state-of-the-art radiative corrections software package: not only model parameters, but also, for instance, which PDF set was used, or which central merging scale values were considered.

One should also  note that the disappearing track analysis (as any long-lived study) can resolve the different lifetimes, while a positive signal in a prompt study does only inform on the mass scale on a given model, but provides no information on the specific lifetime. 

We stress that the disappearing track signature provides unique opportunity  
for the most sensitive test of DM parameter space if long-lived DM charged partners of DM
occurs in the model. This sensitivity outshines the LHC mono-jet constraints, which even for their HL-LHC have a weaker sensitivity to the parameter space. Being concrete, in this study we have found lower bounds on the charged particle mass of 530 GeV for the nominal lifetime of 0.07 ns ($c \tau \sim$ 2 cm ) in the VTDM model, going beyond the 100 GeV limit set by LEP. In the i2HDM model we have found that for $\tau > 0.4$ ns the 95 \% C.L sets a scale of about 237 GeV for a lifetime of 0.5 ns, while for $\tau = 0.05$ ns the limit degrades to 115 GeV.

\acknowledgments 
While this article was in its final stages before publication CMS has made public a new disappearing track study~\cite{CMS-PAS-EXO-19-010}, and the public reinterpretation material was only made public in July 2020~\cite{1790827}. Given the important differences between ATLAS and CMS in a) their inner tracker configurations and b) the analysis strategy, we leave the recasting of the CMS results for future work. 

We are indebted to Ryu Sawada for his valuable help with understanding the details of the ATLAS analysis. We also would like to thank Giovanna Cottin and Nishita Desai for useful discussions, and Andre Lessa for help with uploading our code to the LLP Recasting Repository, for the useful suggestions to improve the code and for being our first beta-tester.

AB acknowledges partial  support from the STFC grant ST/L000296/1 and Soton-FAPESP grant.
AB also thanks the NExT Institute and   Royal Society International Exchange grant IEC-R2-202018. FR acknowledges  Funding for Postdoctoral research in Southampton University, United Kingdom, CONICYT Grant No. 74180065.  This work has been supported by the Mainz Institute for Theoretical Physics (MITP) of the Cluster of Excellence PRISMA+ (Project ID 39083149). S.P.~gratefully acknowledges funding from the Swedish Research Council, under contract number 2016-05996.


\begin{appendix}
 \section{Object reconstruction}
 \label{app:objects}
 In order to ensure the reproducibility of our results, we explicitly list here the kinematical cuts applied on  jets, electrons and muons following~\cite{Aaboud:2017mpt}. The selection for neutralinos, neutrinos and charginos was performed separately taking the data from the branch \textbf{Particle} in the Delphes output, which takes the information directly from PYTHIA8. Since these objects are crucial for the study, we deemed appropriate to explain their treatment in the main text.
 
\subsection{Jets, electrons, muons and charginos}
In table \ref{reconst1} we show the implemented cuts on the transverse momenta $P_T$, Pseudorapidity $\eta$ and Transverse Energy $E_T$ applied to jets, muons, electrons and charginos in the reconstruction stage.
\begin{table}[ht]
\caption{Reconstruction event cuts. The Transverse Energy is defined as $E_T=\sqrt{m^2 + P_T^2}$. When no cut is applied, we indicate it with "X".} 
\begin{center}
  \begin{tabular}{c|c|c|c}
			\hline
			\hline
	  & Transverse     &  Pseudorapidity & Transverse     \\
  	  & Momentum (GeV) &		     & Energy (GeV)   \\ \hline\hline
  Jet     & $P_T > 20$     &  $|\eta|<2.8$   &   X	      \\ \hline
  Muons   & $P_T > 10$     &  $|\eta|<2.7$   &   X	      \\ \hline
Electrons &     X          &  $|\eta|<2.47$  & $E_T > 10$     \\ \hline 
Charginos & $P_T > 5$      &  $|\eta|<2.5$   &    X            \\  \
  \end{tabular} 
	\end{center}
	\label{reconst1}
\end{table}
In addition to the cuts listed in the table above, we also applied the following criteria when dealing with overlaps between the different objects:
\begin{itemize}
\item If $\Delta R$ between a jet and an electron candidate is less than 0.2, the jet is discarded.
\item If an electron and a jet candidate are found between $0.2<\Delta R <0.4$, the electron is discarded.
\item If a muon and a jet candidate are found between $0.2<\Delta R <0.4$, the muon is discarded.
\item If $\Delta R$ between a jet and a muon candidate is less than 0.2, the jet is discarded.
\end{itemize}


\section{Matching procedure}
\label{app:matching}
\begin{figure}[htb]
\centering
{\includegraphics[width=0.53\textwidth]{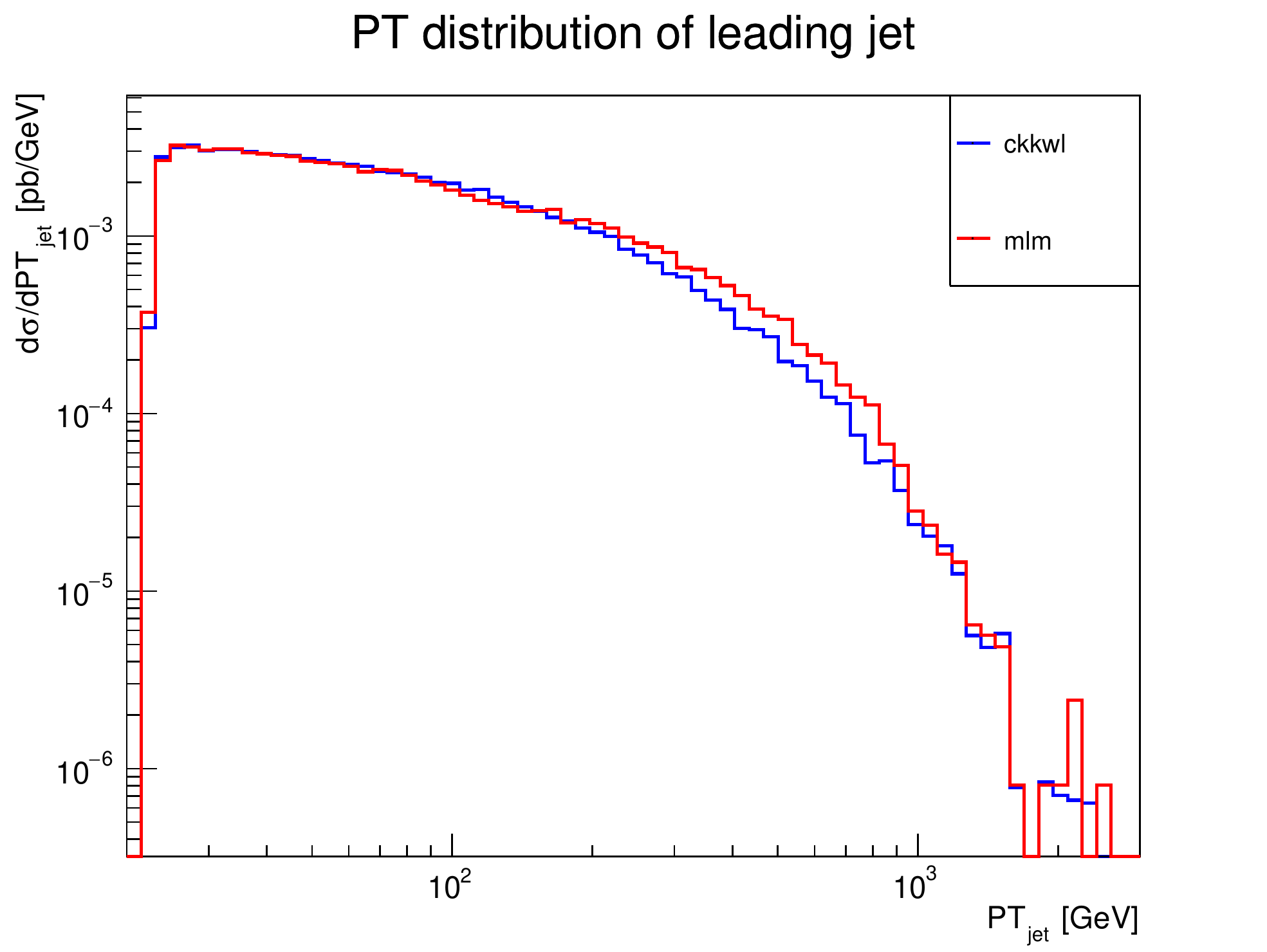}}%
{\includegraphics[width=0.53\textwidth]{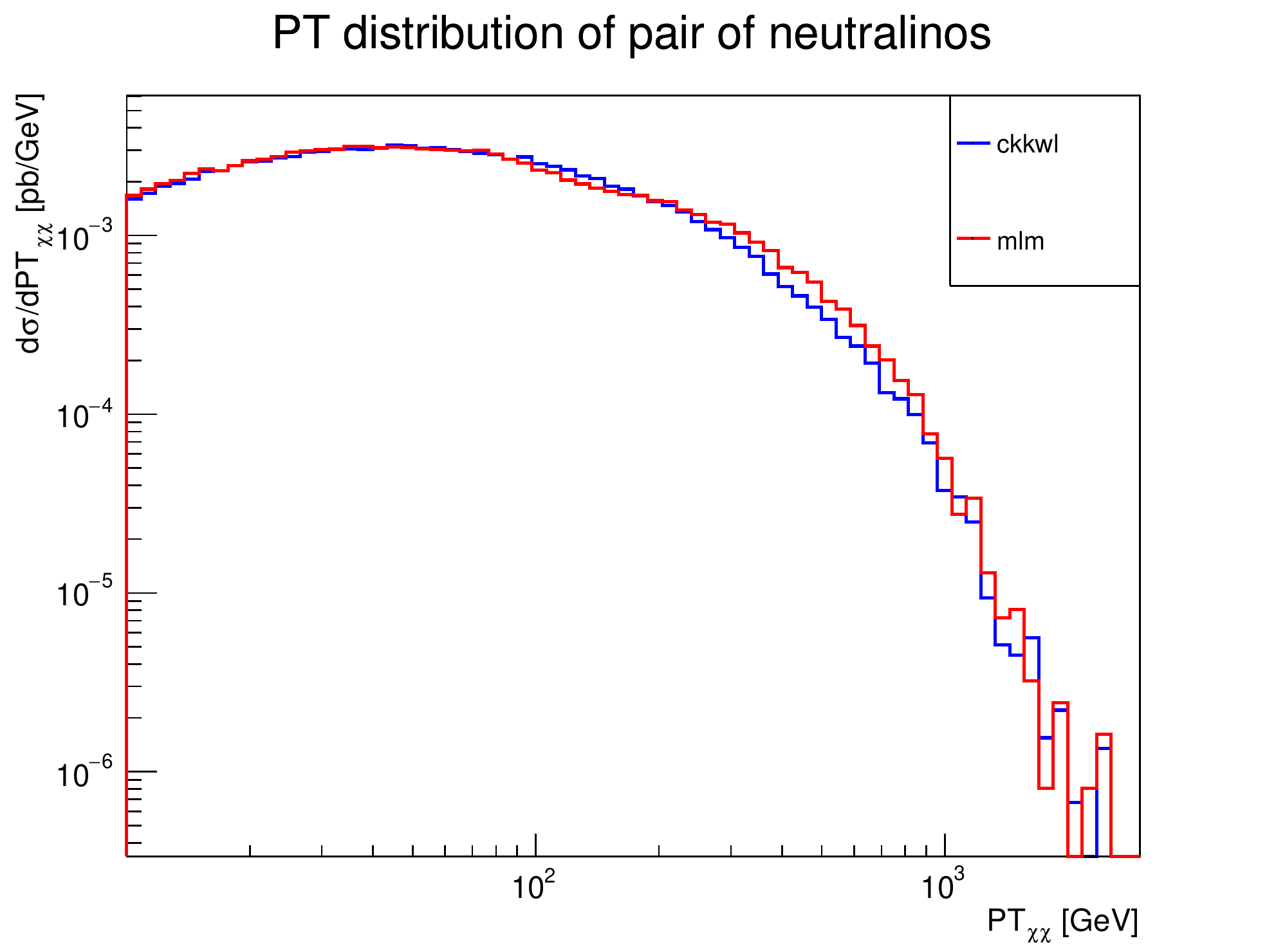}}%
\vskip -0.5cm\hspace*{-3cm}(a)\hspace*{0.48\textwidth}(b)
\caption{Transverse momenta distribution $p_T$ of a) the leading jet and b) pair of neutralinos using CKKW-L (blue) and MLM (red) matching schemes. 
	In this example  $M_{\chi^{\pm}}=400$ GeV, and the  merging scale parameter was set to 100 GeV. \label{fig:matching}}
\end{figure}

In figure~\ref{fig:matching}, we present the distribution of the transverse momentum of the leading jet and a pair of neutralinos using 2 matching schemes: CKKW-L (blue) and MLM (red), using a chargino mass of $M_{\chi^{\pm}}=400$ GeV and a merging scale of $t_{ms}=100$ GeV. Samples with zero and one additional jet at matrix element level are employed. 

At both small and large transverse momenta, the distributions converge, as expected by the requirement that tree-level results for the inclusive production and production in association with one jet should be recovered. However, in the rather extended transition region, visible differences appear. The fist important difference we can observe in both plots is that the MLM scheme present a small valley just after the merging scale at 100 GeV showing the mismatch between the zero- and one-jet sample. This effect is related to details of the jet matching method. The overlap removal strategy by MLM jet matching induces a non-negligible matching scale dependence, which is not analytically tractable, and thus difficult to predict. ``Tuning'' of the matching scale, e.g.\ based on previous experience and reference predictions, can help ameliorate this problem. Similar comments apply for mismatches of renormalization and factorization scales used at fixed order and in the parton shower.

These mismatches are, in principle, absent in the CKKW-L scheme. Indeed, we observe a much smoother transition at the merging scale, for all merging scale values. This leads us to favor CKKW-L for predictions. However, comparison to MLM results can be very informative, by allowing to identify the transition region. 
The absence of mismatches in scale setting in CKKW-L is due to the fact that the parton-shower scale setting procedure is employed throughout phase space, i.e. fixed-order results are reweighted to implement the scale setting. Since the parton-shower scale setting is based on interpreting phase-space points as sequence of single emissions, the resulting scales can lead to unrealistic results for phase-space points that do not follow a simple parton-shower interpretation. In this case, more intricate mechanisms have to be considered~\cite{Fischer:2017yja}. For the present study, the impact of such ``unordered states" is not crucial, as e.g.\ shown by comparison with results of the MLM scheme.

\begin{table}[ht]
\begin{center}
  \begin{tabular}{c|c c| c c| c c| c c}
    \multicolumn{9}{c}{ATLAS benchmark points} \\ \hline \hline
      Signal                      &            \multicolumn{4}{c|}{Event}              &       \multicolumn{4}{c}{Tracklet}                     \\ \hline
                                          & \multicolumn{2}{c|}{$E_A$}&\multicolumn{2}{c|}{$E_E$}&\multicolumn{2}{c|}{$T_A$}&\multicolumn{2}{c}{$T_E$}  \\  \hline
  $m_{\chi}^{\pm}$ (GeV)                  &  ckkwl   &  mlm    & ckkwl &  mlm    & ckkwl &  mlm & ckkwl &  mlm            \\ \hline
            $400$                         &  0.0879  & 0.0815  & 1.030 &  1.021  & 0.08  & 0.08 & 0.44  & 0.44             \\
            $600$                         &  0.1025  & 0.0996  & 1.054 &  1.035  & 0.06  & 0.06 & 0.44  & 0.44             \\
            $600$                         &  0.1025  & 0.0996  & 1.054 &  1.035  & 0.22  & 0.22 & 0.44  & 0.44             \\ \hline
  \end{tabular}
 \end{center}
\caption{Event and tracklet acceptances and efficiencies for the three benchmarks points used by ATLAS, when employing CKKW-L merging or MLM matching.}
\label{tab:accep_match_comparison}
\end{table}

Given the complex selection criteria for the event and tracklet selections, it is not obvious how relevant the differences between matching schemes are in practise. Table~\ref{tab:accep_match_comparison} lists acceptances and efficiencies for two matching schemes. Overall, matching scheme differences have larger impact for low $m_{\chi}^{\pm}$ masses, and for ``more inclusive" \emph{event} selection criteria, reaching up to $7.3\%$ for $E_A$. The acceptance under these criteria include contributions from regions of moderate as well as large jet separation. The dependence on the moderate-separation transition region is removed by the more aggressive tracklet selection. This selection effectively only relies on an accurate model of configurations with two well-separated jets. There is little matching scheme variation in this region, as can be seen in the right-hand edges of both jet spectra in Fig.~\ref{fig:matching}.

\section{Upper Limits and Efficiencies}
\label{app:limits}
In this section, we show tables with efficiencies and excluded regions presented as upper limits on the 2 body production cross section (in fb) in the lifetime (in nanoseconds) - dark matter mass (in GeV) plane for the four different DM Models studied previously. The cross section limits considered the unweighted sume of the channels: $pp\rightarrow D^+D^-$ and $pp\rightarrow D^{\pm}D^0$ and Next-to-leading-Order QCD corrections (k-factor) only for the fermion model. One should note that for small lifetimes, the efficiencies for the various models can differ by an order of magnitude. This is due to the fact that for small lifetimes large boosts are required, which strongly depend on the $p_T$ distributions of the $D^{\pm}$ particles. As shown in figure~\ref{fig:models_xs}d), these distributions differ considerably among the different models studied here.

{\small
\begin{table}[htb]
\begin{center}
\begin{tabular}{l|ccccccc}
\hline \hline
tau & \multicolumn{7}{c}{Mass (GeV)} \\
\cline{2-8} 
(ns)   &   100      &       200 &       300 &       400 &       500 &       600 &       700  \\
\hline
0.01  &  1.37e-06 &  1.90e-07 &  5.64e-08 &  1.86e-08 &  1.17e-08 &  2.59e-11 &  2.41e-09 \\
0.02  &  2.31e-05 &  9.19e-06 &  4.13e-06 &  2.26e-06 &  1.46e-06 &  6.29e-07 &  3.84e-07 \\
0.03  &  8.67e-05 &  5.20e-05 &  3.10e-05 &  2.06e-05 &  1.43e-05 &  8.99e-06 &  6.72e-06 \\
0.04  &  1.90e-04 &  1.43e-04 &  1.02e-04 &  7.52e-05 &  5.61e-05 &  4.06e-05 &  3.24e-05 \\
0.05  &  3.19e-04 &  2.83e-04 &  2.27e-04 &  1.77e-04 &  1.42e-04 &  1.10e-04 &  9.33e-05 \\
0.06  &  4.63e-04 &  4.60e-04 &  4.00e-04 &  3.34e-04 &  2.78e-04 &  2.25e-04 &  1.98e-04 \\
0.07  &  6.14e-04 &  6.67e-04 &  6.15e-04 &  5.32e-04 &  4.61e-04 &  3.89e-04 &  3.47e-04 \\
0.08  &  7.61e-04 &  8.84e-04 &  8.60e-04 &  7.71e-04 &  6.84e-04 &  5.94e-04 &  5.45e-04 \\
0.09  &  9.01e-04 &  1.11e-03 &  1.12e-03 &  1.04e-03 &  9.44e-04 &  8.40e-04 &  7.79e-04 \\
0.10  &  1.03e-03 &  1.35e-03 &  1.40e-03 &  1.32e-03 &  1.23e-03 &  1.12e-03 &  1.05e-03 \\
0.20  &  1.83e-03 &  3.18e-03 &  3.92e-03 &  4.23e-03 &  4.37e-03 &  4.37e-03 &  4.40e-03 \\
0.30  &  2.05e-03 &  4.07e-03 &  5.39e-03 &  6.12e-03 &  6.63e-03 &  6.91e-03 &  7.12e-03 \\
0.40  &  2.05e-03 &  4.41e-03 &  6.08e-03 &  7.13e-03 &  7.92e-03 &  8.45e-03 &  8.85e-03 \\
0.50  &  1.96e-03 &  4.47e-03 &  6.36e-03 &  7.60e-03 &  8.61e-03 &  9.31e-03 &  9.85e-03 \\
0.60  &  1.86e-03 &  4.41e-03 &  6.41e-03 &  7.77e-03 &  8.90e-03 &  9.73e-03 &  1.04e-02 \\
0.70  &  1.75e-03 &  4.29e-03 &  6.33e-03 &  7.75e-03 &  8.97e-03 &  9.89e-03 &  1.06e-02 \\
0.80  &  1.64e-03 &  4.14e-03 &  6.19e-03 &  7.64e-03 &  8.91e-03 &  9.88e-03 &  1.06e-02 \\
0.90  &  1.55e-03 &  3.98e-03 &  6.01e-03 &  7.47e-03 &  8.76e-03 &  9.78e-03 &  1.05e-02 \\
1.00  &  1.46e-03 &  3.81e-03 &  5.82e-03 &  7.28e-03 &  8.59e-03 &  9.60e-03 &  1.04e-02 \\
2.00  &  9.14e-04 &  2.60e-03 &  4.14e-03 &  5.34e-03 &  6.44e-03 &  7.39e-03 &  8.09e-03 \\
3.00  &  6.61e-04 &  1.93e-03 &  3.14e-03 &  4.10e-03 &  5.00e-03 &  5.76e-03 &  6.35e-03 \\
4.00  &  5.18e-04 &  1.54e-03 &  2.52e-03 &  3.31e-03 &  4.06e-03 &  4.70e-03 &  5.20e-03 \\
5.00  &  4.26e-04 &  1.28e-03 &  2.10e-03 &  2.77e-03 &  3.41e-03 &  3.97e-03 &  4.38e-03 \\
6.00  &  3.62e-04 &  1.09e-03 &  1.80e-03 &  2.39e-03 &  2.94e-03 &  3.43e-03 &  3.79e-03 \\
7.00  &  3.16e-04 &  9.52e-04 &  1.58e-03 &  2.09e-03 &  2.58e-03 &  3.02e-03 &  3.33e-03 \\
8.00  &  2.76e-04 &  8.44e-04 &  1.40e-03 &  1.86e-03 &  2.30e-03 &  2.69e-03 &  2.97e-03 \\
9.00  &  2.49e-04 &  7.61e-04 &  1.27e-03 &  1.68e-03 &  2.08e-03 &  2.43e-03 &  2.69e-03 \\
10.00 &  2.25e-04 &  6.88e-04 &  1.15e-03 &  1.52e-03 &  1.89e-03 &  2.21e-03 &  2.45e-03 \\
\bottomrule
\end{tabular}
\end{center}
\caption{Efficiency table for the i2HDM model.}
\end{table}

\clearpage
\newpage

\begin{table}[htb]
\begin{center}
\begin{tabular}{l|cccccccc}
\hline \hline
tau & \multicolumn{8}{c}{Mass (GeV)} \\
\cline{2-9} 
(ns)   &   91      &       200 &       300 &       400 &       500 &       600 &       700 &       800 \\
\hline
0.01  &  4.02e-07 &  8.58e-08 &  8.04e-09 &  9.59e-09 &  3.50e-09 &  0.00e+00 &  0.00e+00 &  0.00e+00 \\
0.02  &  9.40e-06 &  3.35e-06 &  1.27e-06 &  6.60e-07 &  3.26e-07 &  1.11e-07 &  6.49e-08 &  5.91e-08 \\
0.03  &  3.78e-05 &  1.96e-05 &  9.88e-06 &  5.90e-06 &  4.05e-06 &  1.85e-06 &  1.32e-06 &  9.57e-07 \\
0.04  &  8.89e-05 &  5.77e-05 &  3.60e-05 &  2.40e-05 &  1.74e-05 &  9.92e-06 &  7.70e-06 &  5.85e-06 \\
0.05  &  1.57e-04 &  1.22e-04 &  8.37e-05 &  6.08e-05 &  4.71e-05 &  3.08e-05 &  2.51e-05 &  1.96e-05 \\
0.06  &  2.35e-04 &  2.09e-04 &  1.57e-04 &  1.22e-04 &  9.82e-05 &  6.94e-05 &  5.71e-05 &  4.69e-05 \\
0.07  &  3.20e-04 &  3.15e-04 &  2.54e-04 &  2.07e-04 &  1.72e-04 &  1.30e-04 &  1.10e-04 &  9.14e-05 \\
0.08  &  4.08e-04 &  4.37e-04 &  3.74e-04 &  3.17e-04 &  2.69e-04 &  2.14e-04 &  1.87e-04 &  1.56e-04 \\
0.09  &  4.94e-04 &  5.73e-04 &  5.06e-04 &  4.45e-04 &  3.87e-04 &  3.19e-04 &  2.83e-04 &  2.41e-04 \\
0.10  &  5.78e-04 &  7.11e-04 &  6.55e-04 &  5.90e-04 &  5.23e-04 &  4.47e-04 &  4.04e-04 &  3.46e-04 \\
0.20  &  1.13e-03 &  2.08e-03 &  2.34e-03 &  2.46e-03 &  2.40e-03 &  2.38e-03 &  2.31e-03 &  2.15e-03 \\
0.30  &  1.32e-03 &  2.95e-03 &  3.65e-03 &  4.09e-03 &  4.19e-03 &  4.41e-03 &  4.40e-03 &  4.25e-03 \\
0.40  &  1.35e-03 &  3.42e-03 &  4.47e-03 &  5.23e-03 &  5.50e-03 &  5.97e-03 &  6.07e-03 &  6.01e-03 \\
0.50  &  1.32e-03 &  3.63e-03 &  4.94e-03 &  5.94e-03 &  6.38e-03 &  7.07e-03 &  7.29e-03 &  7.33e-03 \\
0.60  &  1.26e-03 &  3.70e-03 &  5.20e-03 &  6.39e-03 &  6.96e-03 &  7.82e-03 &  8.13e-03 &  8.27e-03 \\
0.70  &  1.20e-03 &  3.70e-03 &  5.31e-03 &  6.63e-03 &  7.30e-03 &  8.32e-03 &  8.70e-03 &  8.94e-03 \\
0.80  &  1.14e-03 &  3.64e-03 &  5.32e-03 &  6.74e-03 &  7.51e-03 &  8.62e-03 &  9.08e-03 &  9.39e-03 \\
0.90  &  1.08e-03 &  3.57e-03 &  5.29e-03 &  6.77e-03 &  7.60e-03 &  8.79e-03 &  9.31e-03 &  9.68e-03 \\
1.00  &  1.02e-03 &  3.47e-03 &  5.22e-03 &  6.74e-03 &  7.62e-03 &  8.86e-03 &  9.44e-03 &  9.85e-03 \\
2.00  &  6.64e-04 &  2.55e-03 &  4.08e-03 &  5.57e-03 &  6.54e-03 &  7.85e-03 &  8.57e-03 &  9.16e-03 \\
3.00  &  4.87e-04 &  1.96e-03 &  3.22e-03 &  4.48e-03 &  5.37e-03 &  6.53e-03 &  7.21e-03 &  7.78e-03 \\
4.00  &  3.85e-04 &  1.59e-03 &  2.64e-03 &  3.73e-03 &  4.50e-03 &  5.52e-03 &  6.13e-03 &  6.65e-03 \\
5.00  &  3.18e-04 &  1.33e-03 &  2.23e-03 &  3.17e-03 &  3.86e-03 &  4.76e-03 &  5.30e-03 &  5.78e-03 \\
6.00  &  2.71e-04 &  1.15e-03 &  1.94e-03 &  2.76e-03 &  3.38e-03 &  4.17e-03 &  4.67e-03 &  5.10e-03 \\
7.00  &  2.38e-04 &  1.01e-03 &  1.70e-03 &  2.45e-03 &  3.00e-03 &  3.72e-03 &  4.16e-03 &  4.55e-03 \\
8.00  &  2.09e-04 &  8.99e-04 &  1.52e-03 &  2.19e-03 &  2.70e-03 &  3.35e-03 &  3.76e-03 &  4.11e-03 \\
9.00  &  1.86e-04 &  8.11e-04 &  1.37e-03 &  1.98e-03 &  2.45e-03 &  3.04e-03 &  3.42e-03 &  3.75e-03 \\
10.00 &  1.70e-04 &  7.39e-04 &  1.26e-03 &  1.82e-03 &  2.25e-03 &  2.79e-03 &  3.14e-03 &  3.45e-03 \\
\bottomrule
\end{tabular}
\end{center}
\caption{Efficiency table for the MFDM model.}
\end{table}

\clearpage
\newpage

\begin{table}[htb]
\begin{center}
\begin{tabular}{l|cccccccc}
\hline \hline
tau & \multicolumn{8}{c}{Mass (GeV)} \\
\cline{2-9} 
(ns)   &   100      &       200 &       300 &       400 &       500 &       600 &       700 &       800 \\
\hline
0.01  &  2.27e-04 &  2.12e-05 &  3.45e-06 &  6.90e-07 &  1.53e-07 &  5.27e-08 &  1.06e-08 &  1.10e-10  \\
0.02  &  1.17e-03 &  2.92e-04 &  8.53e-05 &  3.42e-05 &  1.42e-05 &  7.05e-06 &  3.39e-06 &  1.73e-06  \\
0.03  &  2.22e-03 &  8.57e-04 &  3.39e-04 &  1.77e-04 &  9.47e-05 &  5.58e-05 &  3.10e-05 &  1.94e-05  \\
0.04  &  3.10e-03 &  1.54e-03 &  7.49e-04 &  4.56e-04 &  2.86e-04 &  1.87e-04 &  1.15e-04 &  7.82e-05  \\
0.05  &  3.78e-03 &  2.26e-03 &  1.27e-03 &  8.49e-04 &  5.83e-04 &  4.09e-04 &  2.76e-04 &  1.99e-04  \\
0.06  &  4.31e-03 &  2.93e-03 &  1.82e-03 &  1.31e-03 &  9.68e-04 &  7.19e-04 &  5.14e-04 &  3.87e-04  \\
0.07  &  4.71e-03 &  3.55e-03 &  2.40e-03 &  1.83e-03 &  1.41e-03 &  1.10e-03 &  8.23e-04 &  6.36e-04  \\
0.08  &  5.02e-03 &  4.09e-03 &  2.95e-03 &  2.35e-03 &  1.90e-03 &  1.52e-03 &  1.19e-03 &  9.40e-04  \\
0.09  &  5.23e-03 &  4.57e-03 &  3.48e-03 &  2.88e-03 &  2.40e-03 &  1.99e-03 &  1.60e-03 &  1.29e-03  \\
0.10  &  5.39e-03 &  5.00e-03 &  3.98e-03 &  3.39e-03 &  2.91e-03 &  2.47e-03 &  2.03e-03 &  1.67e-03  \\
0.20  &  5.54e-03 &  6.99e-03 &  7.08e-03 &  7.11e-03 &  7.04e-03 &  6.79e-03 &  6.35e-03 &  5.79e-03  \\
0.30  &  4.98e-03 &  7.23e-03 &  8.09e-03 &  8.70e-03 &  9.13e-03 &  9.27e-03 &  9.15e-03 &  8.73e-03  \\
0.40  &  4.41e-03 &  6.97e-03 &  8.26e-03 &  9.24e-03 &  1.00e-02 &  1.05e-02 &  1.07e-02 &  1.04e-02  \\
0.50  &  3.94e-03 &  6.57e-03 &  8.08e-03 &  9.28e-03 &  1.03e-02 &  1.10e-02 &  1.14e-02 &  1.13e-02  \\
0.60  &  3.55e-03 &  6.14e-03 &  7.78e-03 &  9.09e-03 &  1.02e-02 &  1.11e-02 &  1.16e-02 &  1.17e-02  \\
0.70  &  3.23e-03 &  5.75e-03 &  7.43e-03 &  8.80e-03 &  1.00e-02 &  1.10e-02 &  1.16e-02 &  1.18e-02  \\
0.80  &  2.95e-03 &  5.39e-03 &  7.08e-03 &  8.48e-03 &  9.73e-03 &  1.07e-02 &  1.15e-02 &  1.18e-02  \\
0.90  &  2.72e-03 &  5.07e-03 &  6.74e-03 &  8.14e-03 &  9.42e-03 &  1.05e-02 &  1.13e-02 &  1.16e-02  \\
1.00  &  2.52e-03 &  4.76e-03 &  6.41e-03 &  7.80e-03 &  9.10e-03 &  1.01e-02 &  1.10e-02 &  1.13e-02  \\
2.00  &  1.45e-03 &  2.96e-03 &  4.21e-03 &  5.32e-03 &  6.41e-03 &  7.35e-03 &  8.18e-03 &  8.64e-03  \\
3.00  &  1.02e-03 &  2.14e-03 &  3.11e-03 &  3.98e-03 &  4.86e-03 &  5.64e-03 &  6.34e-03 &  6.76e-03  \\
4.00  &  7.91e-04 &  1.68e-03 &  2.47e-03 &  3.17e-03 &  3.90e-03 &  4.55e-03 &  5.15e-03 &  5.52e-03  \\
5.00  &  6.42e-04 &  1.38e-03 &  2.03e-03 &  2.63e-03 &  3.26e-03 &  3.81e-03 &  4.33e-03 &  4.65e-03  \\
6.00  &  5.42e-04 &  1.17e-03 &  1.73e-03 &  2.26e-03 &  2.80e-03 &  3.28e-03 &  3.73e-03 &  4.02e-03  \\
7.00  &  4.68e-04 &  1.02e-03 &  1.51e-03 &  1.97e-03 &  2.45e-03 &  2.87e-03 &  3.28e-03 &  3.53e-03  \\
8.00  &  4.12e-04 &  8.95e-04 &  1.34e-03 &  1.75e-03 &  2.18e-03 &  2.56e-03 &  2.93e-03 &  3.15e-03  \\
9.00  &  3.69e-04 &  8.05e-04 &  1.20e-03 &  1.58e-03 &  1.95e-03 &  2.31e-03 &  2.63e-03 &  2.85e-03  \\
10.00 &  3.32e-04 &  7.26e-04 &  1.09e-03 &  1.42e-03 &  1.78e-03 &  2.11e-03 &  2.40e-03 &  2.59e-03  \\
\bottomrule
\end{tabular}
\end{center}
\caption{Efficiency table for the VTDM model.}
\end{table}
\clearpage

\newpage
\begin{table}[htb]
\begin{center}
\begin{tabular}{l|cccccccc}
\hline \hline
tau & \multicolumn{7}{c}{Mass (GeV)} \\
\cline{2-8} 
(ns)   &   100      &       200 &       300 &       400 &       500 &       600 &       700  \\
\hline
0.01  &  1.61e+05 &    1.16e+06 &    3.90e+06 &     1.19e+07 &     1.88e+07 & 8.49e+09 &   9.12e+07 \\
0.02  &      9506 &       23930 &       53300 &        97200 &     1.51e+05 & 3.50e+05 &   5.73e+05 \\
0.03  &      2537 &        4230 &        7092 &        10690 &        15340 &  24460 &        32760 \\
0.04  &      1156 &        1533 &        2161 &         2926 &         3924 &   5416 &         6791 \\
0.05  &     689.4 &       778.7 &       968.7 &         1242 &         1554 &   2006 &         2359 \\
0.06  &     475.3 &       478.6 &       549.9 &        659.2 &        792.6 &  975.7 &         1112 \\
0.07  &     358.5 &       330.1 &       357.6 &        413.4 &        476.7 &  565.8 &        633.3 \\
0.08  &     289.3 &       248.9 &       255.7 &        285.2 &        321.9 &  370.6 &        404.0 \\
0.09  &     244.1 &       197.5 &       195.9 &        212.1 &        233.0 &  261.8 &        282.3 \\
0.10  &     213.1 &       163.5 &       157.1 &        166.3 &        178.7 &  196.7 &        209.7 \\
0.20  &     119.9 &       69.09 &       56.13 &        52.06 &        50.38 &  50.31 &        50.03 \\
0.30  &     107.1 &       54.02 &       40.82 &        35.95 &        33.20 &  31.85 &        30.89 \\
0.40  &     107.5 &       49.93 &       36.19 &        30.85 &        27.76 &  26.03 &        24.85 \\
0.50  &     112.2 &       49.16 &       34.59 &        28.95 &        25.56 &  23.64 &        22.34 \\
0.60  &     118.4 &       49.85 &       34.35 &        28.32 &        24.72 &  22.61 &        21.21 \\
0.70  &     125.9 &       51.28 &       34.74 &        28.38 &        24.53 &  22.25 &        20.76 \\
0.80  &     134.0 &       53.16 &       35.54 &        28.78 &        24.69 &  22.27 &        20.69 \\
0.90  &     141.8 &       55.28 &       36.58 &        29.44 &        25.11 &  22.50 &        20.86 \\
1.00  &     150.7 &       57.70 &       37.81 &        30.21 &        25.62 &  22.91 &        21.17 \\
2.00  &     240.7 &       84.71 &       53.11 &        41.19 &        34.15 &  29.78 &        27.19 \\
3.00  &     332.6 &       114.0 &       70.02 &        53.61 &        44.03 &  38.18 &        34.63 \\
4.00  &     425.0 &       143.0 &       87.26 &        66.39 &        54.24 &  46.78 &        42.32 \\
5.00  &     516.6 &       172.4 &       104.6 &        79.31 &        64.56 &  55.40 &        50.21 \\
6.00  &     608.3 &       201.9 &       122.1 &        92.24 &        74.80 &  64.15 &        58.07 \\
7.00  &     697.2 &       231.1 &       139.4 &        105.2 &        85.16 &  72.91 &        66.03 \\
8.00  &     796.2 &       260.7 &       156.7 &        118.1 &        95.52 &  81.71 &        73.98 \\
9.00  &     885.1 &       289.2 &       173.7 &        131.0 &        105.6 &  90.54 &        81.93 \\
10.00 &     978.6 &       319.6 &       191.9 &        144.3 &        116.2 &  99.56 &        89.98 \\
\bottomrule
\end{tabular}
\end{center}
\caption{Upper limits for production cross section in fb for the i2HDM model.}
\end{table}
\clearpage

\newpage
\begin{table}[tb]
\begin{center}
\begin{tabular}{l|cccccccc}
\hline \hline
tau & \multicolumn{8}{c}{Mass (GeV)} \\
\cline{2-9} 
(ns)  &   91      &       200 &       300 &       400 &       500 &       600 &       700 &       800 \\
\hline
0.01  &  5.46e+05 &    2.56e+06 &     2.73e+07 &     2.29e+07 &     6.28e+07 &         --- &         --- &         --- \\
0.02  &  2.34e+04 &    6.56e+04 &     1.74e+05 &     3.34e+05 &     6.75e+05 &    1.98e+06 &    3.39e+06 &    3.72e+06 \\
0.03  &      5815 &    1.12e+04 &        22280 &        37310 &        54380 &      119200 &      166700 &      230000 \\
0.04  &      2474 &        3812 &         6119 &         9178 &        12670 &       22170 &       28580 &       37600 \\
0.05  &      1404 &        1805 &         2628 &         3621 &         4673 &        7152 &        8770 &       11220 \\
0.06  &     937.9 &        1051 &         1404 &         1806 &         2240 &        3172 &        3852 &        4688 \\
0.07  &     687.7 &       698.4 &          865 &         1064 &         1278 &        1689 &        1999 &        2406 \\
0.08  &     539.6 &       503.4 &        588.8 &        694.8 &        818.3 &        1028 &        1179 &        1408 \\
0.09  &     445.8 &       384.1 &        435.1 &        493.9 &        568.5 &       690.4 &       777.3 &       913.7 \\
0.10  &     380.8 &       309.2 &        336.0 &        372.7 &        420.9 &       492.2 &       545.0 &       635.5 \\
0.20  &     194.1 &       106.0 &         94.0 &        89.53 &        91.84 &       92.38 &       95.35 &       102.4 \\
0.30  &     166.9 &       74.47 &        60.23 &        53.76 &        52.55 &       49.92 &       50.02 &       51.71 \\
0.40  &     163.1 &       64.41 &        49.20 &        42.09 &        40.03 &       36.87 &       36.23 &       36.60 \\
0.50  &     167.2 &       60.56 &        44.54 &        37.04 &        34.48 &       31.14 &       30.18 &       30.01 \\
0.60  &     174.7 &       59.39 &        42.30 &        34.45 &        31.62 &       28.15 &       27.05 &       26.59 \\
0.70  &     183.9 &       59.54 &        41.45 &        33.18 &        30.12 &       26.46 &       25.27 &       24.61 \\
0.80  &     193.8 &       60.39 &        41.33 &        32.64 &        29.31 &       25.53 &       24.24 &       23.43 \\
0.90  &     204.1 &       61.70 &        41.61 &        32.50 &        28.95 &       25.03 &       23.62 &       22.73 \\
1.00  &     215.3 &       63.43 &        42.12 &        32.63 &        28.87 &       24.83 &       23.31 &       22.32 \\
2.00  &     331.5 &       86.40 &        53.87 &        39.53 &        33.61 &       28.02 &       25.67 &       24.00 \\
3.00  &     451.7 &      112.10 &        68.31 &        49.10 &        40.96 &       33.67 &       30.53 &       28.27 \\
4.00  &     570.9 &      138.70 &        83.31 &        59.05 &        48.86 &       39.84 &       35.87 &       33.10 \\
5.00  &     692.0 &      165.20 &        98.62 &        69.35 &        56.94 &       46.24 &       41.48 &       38.09 \\
6.00  &     812.4 &      191.20 &       113.70 &        79.58 &        65.15 &       52.71 &       47.09 &       43.17 \\
7.00  &     925.7 &      217.70 &       129.20 &        89.98 &        73.29 &       59.20 &       52.86 &       48.30 \\
8.00  &    1051.0 &      244.70 &       144.70 &       100.60 &        81.42 &       65.69 &       58.58 &       53.48 \\
9.00  &    1182.0 &      271.20 &       160.20 &       111.10 &        89.71 &       72.31 &       64.33 &       58.61 \\
10.00 &    1291.0 &      297.90 &       175.30 &       121.00 &        98.00 &       78.80 &       70.09 &       63.85 \\
\bottomrule
\end{tabular}
\end{center}
\caption{Upper limits for production cross section in fb for the MFDM model. Cells with entry "---" contain an infinite upper limit (i.e. no sensitivity). This is due to a vanishing efficiency in the corresponding cell of the efficiency table, which in turn is a consequence of the low statistic of the heatmap used to describe the analysis.}
\end{table}

\clearpage
\newpage
\begin{table}[tb]
\begin{center}
\begin{tabular}{l|cccccccc}
\hline \hline
tau & \multicolumn{8}{c}{Mass (GeV)} \\
\cline{2-9} 
(ns)  &   91      &       200 &       300 &       400 &       500 &       600 &       700 &       800 \\
\hline
0.01  &   968.4 &     10390 &     63800 &     318700 &    1.44e+06 &    4.17e+06 &     2.08e+07 &  1.993e+09 \\
0.02  &   187.4 &     753.3 &      2580 &       6434 &       15530 &       31210 &        64850 &  1.272e+05 \\
0.03  &   99.06 &     256.7 &     649.0 &       1246 &        2324 &        3940 &         7094 &  11360     \\
0.04  &   70.91 &     142.5 &     293.7 &      482.5 &       768.2 &        1179 &         1909 &   2814     \\
0.05  &   58.26 &     97.35 &     173.7 &      259.1 &       377.6 &         538 &        797.9 &   1107     \\
0.06  &   51.03 &     74.99 &     120.8 &      167.5 &       227.3 &       305.9 &        427.8 &  568.8     \\
0.07  &   46.74 &     61.99 &     91.81 &      120.3 &       155.8 &       200.1 &        267.2 &  346.2     \\
0.08  &   43.78 &     53.73 &     74.67 &       93.6 &       115.7 &       144.3 &        185.5 &  234.1     \\
0.09  &   42.08 &     48.14 &     63.26 &      76.32 &       91.49 &       110.7 &        137.8 &  171.0     \\
0.10  &   40.82 &     44.04 &     55.33 &      64.85 &       75.57 &       89.13 &        108.5 &  131.5     \\
0.20  &   39.73 &     31.48 &     31.05 &      30.94 &       31.24 &       32.42 &        34.65 &  38.01     \\
0.30  &   44.16 &     30.43 &     27.20 &      25.28 &       24.10 &       23.72 &        24.04 &  25.21     \\
0.40  &   49.87 &     31.55 &     26.64 &      23.81 &       21.96 &       21.04 &        20.65 &  21.10     \\
0.50  &   55.79 &     33.48 &     27.21 &      23.70 &       21.43 &       20.07 &        19.35 &  19.46     \\
0.60  &   62.01 &     35.80 &     28.28 &      24.20 &       21.52 &       19.89 &        18.90 &  18.79     \\
0.70  &   68.19 &     38.24 &     29.61 &      24.99 &       21.96 &       20.08 &        18.89 &  18.63     \\
0.80  &   74.55 &     40.82 &     31.07 &      25.93 &       22.60 &       20.47 &        19.12 &  18.72     \\
0.90  &   80.84 &     43.43 &     32.66 &      27.04 &       23.36 &       21.04 &        19.51 &  18.99     \\
1.00  &   87.22 &     46.19 &     34.35 &      28.19 &       24.18 &       21.68 &        20.00 &  19.39     \\
2.00  &   151.5 &     74.27 &     52.21 &      41.35 &       34.33 &       29.92 &        26.88 &  25.45     \\
3.00  &   214.7 &     102.9 &     70.72 &      55.27 &       45.31 &       38.97 &        34.71 &  32.56     \\
4.00  &   278.2 &     131.0 &     89.24 &      69.35 &       56.41 &       48.31 &        42.70 &  39.88     \\
5.00  &   342.8 &     159.6 &     108.3 &      83.51 &       67.45 &       57.69 &        50.81 &  47.35     \\
6.00  &   405.8 &     188.5 &     127.1 &      97.46 &       78.65 &       66.98 &        58.93 &  54.77     \\
7.00  &   470.1 &     215.9 &     145.6 &      111.7 &       89.95 &       76.58 &        67.03 &  62.31     \\
8.00  &   534.0 &     245.8 &     164.0 &      126.1 &      101.00 &       85.79 &        75.19 &  69.79     \\
9.00  &   596.9 &     273.2 &     182.9 &      139.7 &      112.50 &       95.21 &        83.59 &  77.21     \\
10.00 &   662.8 &     303.2 &     201.7 &      154.7 &      123.40 &      104.40 &        91.62 &  84.86     \\
\bottomrule
\end{tabular}
\end{center}
\caption{Upper limits for production cross section in fb for the VTDM model.}
\end{table}
\clearpage
}

\end{appendix}

\newpage
\bibliographystyle{JHEP}
\bibliography{LLP_DM_paper}

\end{document}